\documentclass[journal]{IEEEtran}
\usepackage{amsfonts}
\usepackage{mathrsfs}
\usepackage{amssymb,amsmath}
\usepackage[noblocks]{authblk}
\usepackage{stfloats}
\usepackage{algorithm}
\usepackage{algorithmic}
\usepackage{amsmath,amssymb,epsfig,graphics,subfigure}
\usepackage{theorem}
\usepackage{array,color}
\usepackage[compress]{cite}
\usepackage{subfigure}
\usepackage{url}
\usepackage{multirow}

\newtheorem{lemma}{Lemma}

\theoremheaderfont{\normalfont\bfseries}

\begin{document}
\title{Session-Based Cooperation in Cognitive Radio Networks: A Network-Level Approach}
\author{\IEEEauthorblockN{Haichuan Ding, Chi Zhang,~\IEEEmembership{Member,~IEEE}, Xuanheng Li,~\IEEEmembership{Student Member,~IEEE}, Jianqing Liu,~\IEEEmembership{Student Member,~IEEE}, Miao Pan,~\IEEEmembership{Member,~IEEE}, Yuguang Fang,~\IEEEmembership{Fellow,~IEEE} and Shigang Chen,~\IEEEmembership{Fellow,~IEEE}
\thanks{This work was partially supported by the U.S. National Science Foundation under grants CNS-1343356 and CNS-1409797.}
\thanks{Haichuan Ding, Jianqing Liu and Yuguang Fang are with the Department of Electrical and Computer Engineering, University of Florida, Gainesville, FL 32611, USA (email: dhcbit@gmail.com, jianqingliu@ufl.edu, fang@ece.ufl.edu).}
\thanks{Chi Zhang is with the School of Information Science and Technology, University of Science and Technology of China, Hefei 230027, China (email: chizhang@ustc.edu.cn).}
\thanks{Xuanheng Li is with the School of Information and Communication Engineering, Dalian University of Technology, Dalian, China 116023 (email: lixuanheng@mail.dlut.edu.cn).}
\thanks{Miao Pan is with the Department of Electrical and Computer Engineering, University of Houston, Houston, TX 77004, USA (email: mpan2@uh.edu).}
\thanks{Shigang Chen is with the Department of Computer and Information Science and Engineering, University of Florida, Gainesville, FL 32611 USA (e-mail: sgchen@cise.ufl.edu).}
}}

\maketitle
\begin{abstract}
In cognitive radio networks (CRNs), secondary users (SUs) can proactively obtain spectrum access opportunities by helping with primary users' (PUs') data transmissions. Currently, such kind of spectrum access is implemented via a {\em cooperative communications based link-level frame-based cooperative (LLC)} approach where {\em individual} SUs {\em independently} serve as relays for PUs in order to gain spectrum access opportunities. Unfortunately, this LLC approach cannot fully exploit spectrum access opportunities to enhance the throughput of CRNs and fails to motivate PUs to join the spectrum sharing processes. To address these challenges, we propose a {\em network-level session-based cooperative (NLC)} approach where SUs are {\em grouped together} to cooperate with PUs {\em session by session}, instead of frame by frame as what has been done in existing works, for spectrum access opportunities of the corresponding group. Thanks to our group-based session-by-session cooperating strategy, our NLC approach is able to address all those challenges in the LLC approach. To articulate our NLC approach, we further develop an NLC scheme under a cognitive capacity harvesting network (CCHN) architecture. We formulate the cooperative mechanism design as a cross-layer optimization problem with constraints on primary session selection, flow routing and link scheduling. To search for solutions to the optimization problem, we propose an augmented scheduling index ordering based (SIO-based) algorithm to identify maximal independent sets. Through extensive simulations, we demonstrate the effectiveness of the proposed NLC approach and the superiority of the augmented SIO-based algorithm over the traditional method.
\end{abstract}


\section{Introduction}
Ever since the Federal Communications Commission (FCC) opened up opportunities for unlicensed users to access under-utilized licensed spectrum bands to address the spectrum crisis, considerable research efforts have been devoted to enabling dynamic spectrum access, with cognitive radio (CR) technology, which potentially offers a promising
solution \cite{Cisco2016,Bazelon2015,Chen2016,McHenry2006,Haykin2005}. Cognitive radios allow secondary users (SUs) to proactively help with primary users' (PUs') transmissions in order to gain spectrum access opportunities as a reward \cite{Haykin2005,Zhang20141,Ren2016,Simeone2008}. In the subsequent development, this spectrum access paradigm will be referred to as the cooperation-based spectrum access for simplicity since its effectiveness relies on the cooperation between PUs and SUs.

In the current literature, the cooperation-based spectrum access is implemented through a link-level frame-based cooperative (LLC) approach which is built on cooperative communications in physical layer. In the LLC approach, PUs employ SUs as relays to expedite data transmission for each MAC frame so that the saved frame transmission time can be offered to SUs for spectrum access, and individual SUs independently cooperate with PUs for their
own spectrum access opportunities \cite{Simeone2008,Cao2013,Jing2015,Duan2014,Feng2014,Li2013,Zhang2013}. Although the LLC approach is able to maximize the achievable throughput of relay SUs, it cannot efficiently exploit available spectrum access opportunities in cognitive radio networks (CRNs) to improve network-level throughput.

\begin{figure}[!t]
 \begin{center}
  \includegraphics[width=2.5in]{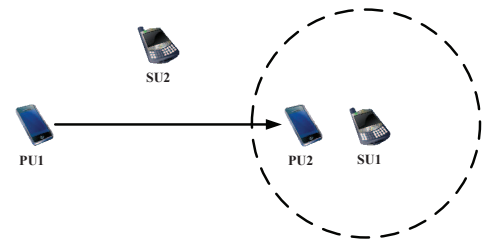}
  \end{center}
  \begin{center}
   \parbox{8cm}{\caption{The employed LLC scheme may lead to inefficient resource utilization. The dashed circle signifies the interference range of SU1.}  \label{fig}}
  \end{center}
  \end{figure}

An underlying assumption in the LLC approach is that individual SUs independently cooperate with PUs for their own spectrum access opportunities and the generated spectrum access opportunities are exclusively granted to relaying SUs such that other SUs cannot transmit during the cooperation-incurred periods \cite{Long2014,Zhang2014}. As a result, the LLC approach will miss a significant number of spectrum access opportunities to improve the throughput of CRNs. This is illustrated by the example shown in Fig. \ref{fig} where SU1 want to access PUs' spectrum for data transmissions while PU1 is transmitting a file to PU2. SU2 does not have data to transmit. Because of unfavorable position, SU1 is unable to cooperate with PUs to gain spectrum access opportunities while SU2 is able to do so. In this case, if SU2 is willing to cooperate with PUs to acquire spectrum access opportunities and offer these opportunities to SU1, SU1 will be able to transmit its data and the throughput of the CRN is improved. Unfortunately, this is not supported by the LLC approach where SU1 and SU2 independently cooperates with PUs for their own spectrum access opportunities. According to the LLC approach, SU2 will not cooperate with PUs since it does not have data to transmit. As a result, the LLC approach is unable to fully exploit available spectrum access opportunities in CRNs for network-level throughput enhancement.

Motivated by this observation, in this paper, we propose a network-level session-based cooperative (NLC) approach for CRNs so that the spectrum access opportunities are utilized more efficiently. Unlike the LLC approach where SUs cooperate with PUs for {\em their own spectrum access opportunities}, in our NLC approach, SUs are grouped together and cooperate with PUs for {\em the spectrum access opportunities of the corresponding group}. In our NLC approach, SUs need to share obtained spectrum access opportunities among a group of SUs according to certain strategies such that other SUs are able to obtain spectrum access opportunities despite their unfavorable locations. For the example in Fig. \ref{fig}, under our NLC approach, SU2 first cooperates with PUs for spectrum access opportunities and, then, shares its spectrum access opportunities with SU1. Thus, SU1 will be able to access PUs' spectrum for data transmission and the capacity of the considered CRN will be improved. Specifically, in our approach, SUs, as a group, first help expedite data transmission for PUs' sessions, namely, primary sessions, so that the scheduled data of corresponding primary sessions are delivered in shorter time periods than what are scheduled, e.g., two-third of the scheduled time, and, then, the remaining time of the primary sessions are granted to this group of SUs for spectrum access.

\begin{figure}[!t]
 \begin{center}
  \includegraphics[width=2.2in]{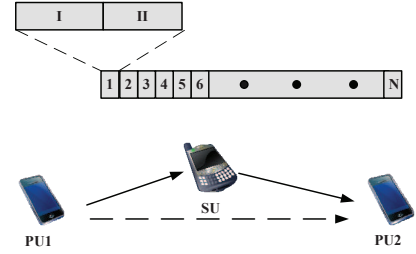}
  \end{center}
  \begin{center}
   \parbox{8cm}{\caption{SU intends to access the spectrum allocated to PU1 and PU2 while PU1 is delivering a file to PU2. The file is expected to be deliver in $N$ ($N>>1$) frames without the help of SU. Those digits in this figure represent different frames. The Roman numerals signify the division of each frame in the cooperative communication based LLC approach.}  \label{fig1}}
  \end{center}
  \end{figure}

Another salient feature of our NLC approach, when compared with the LLC approach, is that our approach works session by session instead of frame by frame. To further elaborate this difference between the LLC approach and our NLC approach, let us consider the example shown in Fig. \ref{fig1} where PU1 wants to transmit a file to PU2 in $N$ $(N>>1)$ MAC frames. Following the LLC approach, each frame is divided into two parts which are indicated by the Roman numerals. In the first part of a frame, SU helps with PU1's data transmission via a cooperative communication scheme in physical layer, such as decode-and-forward or amplify-and-forward relaying, so that PU1's scheduled data is delivered before its intended time without cooperation. Thus, PU1 grants the spectrum access opportunities in the second part of the frame to SU as a reward, whereas our NLC approach requires SU to help PU1 deliver the whole file to PU2 in, for example, $\frac{2}{3}N$ frames and obtain the remaining $\frac{1}{3}N$ frames for its own data transmissions. Clearly, the feasibility of the cooperation-based spectrum access highly relies on PUs' willingness to yield certain spectrum access opportunities in exchange for reduced latency of service delivery. Unfortunately, the LLC approach might fail to benefit PUs and thus might not be able to enable the cooperation-based spectrum access. As shown in Fig. \ref{fig1}, even if SU can help expedite PU1's file transferring process in each frame by following the LLC approach, PU1 still needs to wait until the last frame for the whole file to be delivered. As a result, PU1 might not actually benefit from SU's help and thus might not be interested in cooperating with SU \cite{Garcia2004}. In contrast to the LLC approach where SUs look for cooperative communications based spectrum access opportunities frame by frame, our NLC approach works session by session and requires SUs to help with PUs' end-to-end (E2E) data delivery in exchange for spectrum access opportunities. Hence, PUs will benefit from our NLC approach since the latency of their E2E service delivery will be significantly improved with the help of SUs and thus will be willing to yield spectrum access opportunities to SUs in exchange for enhanced quality of service.

As mentioned above, our NLC approach only provides a way for spectrum sharing and its nice features cannot be efficiently exploited without a suitable network architecture. To facilitate our NLC approach, necessary control messages, such as those for spectrum sharing, must be exchanged among SUs so that their actions are well coordinated. In literature, this is often achieved through the common control channels (CCCs) \cite{Thilina2015,Anamalamudi2016}. Unfortunately, when SUs seek for opportunistic access to PUs' spectrum, they are likely already short of available spectrum resources for information exchange and do not have extra resources for CCC establishment \cite{Thilina2015,Zhang20141}. In view of this as well as the selfishness of SUs, it is difficult for SUs to enjoy the benefits promised by the NLC approach without a network-level solution. To fully exploit the benefits of our NLC approach, in this paper, we develop an NLC scheme for CRNs under a cognitive capacity harvesting network (CCHN) architecture where a secondary service provider (SSP) deploys base stations (BSs) and cognitive radio routers (CR routers) to provide secondary services to SUs\footnote{The SSP is an independent service provider which has its own licensed bands, such as the cellular service provider. BSs deployed by the SSP is to provide the support of basic reliable service, just as in typical cellular systems. BSs are connected to data networks via wired connections and serve as gateways for CR-routers. BSs also work as an agent for the SSP to exchange control signaling with CR routers and SUs. CR routers are wireless routers with multiple interfaces and can support multiple communication technologies. Detailed introduction to entities in the CCHN will be provided in Section II.} \cite{Pan2012,Grndalen2011,Cammarano2015,Fodor2009}. In our NLC scheme, individual SUs only need to access the SSP's network, i.e., the CCHN, for services. It is the SSP and its deployed infrastructure that cooperate with PUs to gain spectrum access opportunities. This design frees SUs from the cooperating process and thus reduces user-side complexity. Under the supervision of the SSP, BSs and CR routers, as a group, cooperate with PUs to gain spectrum access opportunities for the CCHN. After that, the obtained spectrum access opportunities are efficiently allocated among those BSs and CR routers by the SSP to enhance network capacity. Specifically, in our NLC scheme, the SSP obtains PUs' traffic information, such as session lengths\footnote{The length of a primary session is the duration of the intended transmission periods of this session.} and data volumes, from the primary service provider (PSP) and coordinates BSs and CR routers to cooperate via its own spectrum bands, i.e., the SSP's basic bands\footnote{The basic bands considered here are mainly used to deliver control messages among BSs and CR routers although it can be used to provide other services if available.}. Once PUs' data is delivered, the remaining time of those primary sessions are granted to the CCHN, i.e., the SSP, for spectrum access and the SSP allocates the cooperation-incurred transmission opportunities among BSs and CR routers to serve SUs.
Under the CCHN, we demonstrate the feasibility of the NLC scheme as well as the impact of various network parameters through a throughput maximization problem. Our major contributions are summarized as follows:

\begin{itemize}
  \item This is the first work to consider network-level session-based cooperation for CRNs. Unlike existing schemes, the proposed NLC scheme is a network-wide cooperative scheme where BSs and CR routers deployed by the SSP, as a group, cooperate with PUs for spectrum access opportunities of the CCHN. In order to motivate PUs to join the cooperation-based spectrum sharing processes, our NLC scheme focuses on E2E service provisioning for PUs and works with sessions instead of MAC frames, which further differentiates our NLC scheme from existing ones. As an independent service provider, the SSP optimally selects BSs and CR routers to cooperate with PUs and intelligently allocates cooperation-incurred periods among BSs and CR routers to establish multi-hop connections to serve SUs, resulting in efficient utilization of the cooperation-incurred spectrum access opportunities.
  \item To characterize interference relations in the CCHN, we introduce a PU-related conflict graph. Different from the existing work, the PU-related conflict graph not only characterizes conflicting relationship between CR links\footnote{CR links refer to links between BSs, links between CR routers, and links between BSs and CR routers.}, but also captures conflicts among CR links, PU-related links\footnote{PU-related links refer to links from sources of primary sessions to BSs or CR routers and links from BSs or CR routers to destinations of primary sessions.}, and primary sessions. The maximal independent sets (MISs) of the PU-related conflict graph are critical for later establishment of the cross-layer constraints. Yet searching for all MISs in the PU-related conflict graph is NP-complete. We develop an augmented scheduling index ordering based (SIO-based) algorithm to address this problem based on the SIO method from \cite{Li2010}.
  \item We formulate the cooperative mechanism design as a cross-layer optimization problem to maximize the throughput of the CCHN by jointly considering primary session selection, flow routing and link scheduling constraints. To capture the incentives for PUs, we introduce an incentive parameter $\alpha$ ($\alpha \ge 1$) to indicate that PUs are willing to cooperate with the SSP once the SSP promises to deliver their data in $\frac{1}{\alpha}$ of the scheduled times. Noticing that the optimal achievable throughput of the CCHN is affected by the SSP's cooperating decisions on primary sessions, we include primary session selection and primary flow routing constraints in our problem formulation, which further differentiates our work from existing ones. In contrast to the case with fixed frame length at the physical/MAC layer, our formulation allows differentiation in session lengths and data volumes of primary sessions.
  \item By carrying out extensive simulations, we demonstrate the feasibility of the proposed NLC scheme and study the impact of various network parameters. Additionally, we examine the performance of the proposed MIS-searching algorithm, and the results confirm the superiority of the augmented SIO-based algorithm (proposed in this paper) over the original SIO-based method.
\end{itemize}

For the readers' convenience, the important notations used in this paper is summarized in Table \ref{tab}.

The rest of the paper is organized as follows. Related works are reviewed in Section II. In Section III, we give a detailed description of the CCHN and the NLC scheme. Then, we present corresponding network settings and related models in Section IV. In Section V, we formulate the cooperative mechanism design as an optimization problem and present the augmented SIO-based algorithm for critical MISs searching. Performance evaluation results and corresponding analysis are presented in Section VI. Finally, the achievable throughput of individual SUs are discussed in Section VII, and concluding remarks are drawn in Section VIII.

\begin{table}[!hbp]
\begin{center}
\caption{List of Important Notations and Definitions.}
\label{tab}
\begin{tabular}{|c|c|}
\hline
Notation & Definition\\
\hline
$\mathcal{N}$& The index set of CR routers\\
\hline
$\mathcal{L}$ & The index set of edge CR routers\\
\hline
$s\left(l\right)$& The $l$th edge CR router\\
\hline
$\mathcal{L}_p$& The index set of primary sessions\\
\hline
$s\left(l_p\right)$ & The source of the $l_p$th primary session\\
\hline
$d\left(l_p\right)$& The destination of the $l_p$th primary session\\
\hline
$T_{l_p}$ & The length of the $l_p$th primary session\\
\hline
$D_{l_p}$ & The data volume of the $l_p$th primary session\\
\hline
\multirow{2}{*}{$\theta_{l_p}$} & =1 if the SSP chooses to cooperate with \\
&the $l_p$th primary session and is $0$ otherwise\\
\hline
\multirow{2}{*}{$G=\left(V,E\right)$} & $G$ is the PU-related conflict graph, \\
&$V$ is the vertex set of $G$, and $E$ is the edge set of $G$\\
\hline
$\mathcal{I}$ & The set of maximal independent sets (MISs) of $G$\\
\hline
\multirow{2}{*}{$\mathcal{I}_{l_p}$} & The set of MISs containing the vertex corresponding\\
&to the $l_p$th primary session\\
\hline
\multirow{2}{*}{$\overline{\mathcal{I}}_{l_p}$}& A set of MISs, satisfying ${\mathcal{I}_{{l_p}}}  \cap \overline{\mathcal{I}}_{{l_p}}=\emptyset$\\
&and ${\mathcal{I}_{{l_p}}}  \cup \overline{\mathcal{I}}_{{l_p}}=\mathcal{I}$\\
\hline
$T$ & Control Interval for the SSP\\
\hline
\multirow{2}{*}{$f_{ij}(l)$} & The amount of the $l$th secondary flow allocated on\\ &the link between the $i$th and the $j$th CR routers\\
\hline
\multirow{2}{*}{$f_{ij}^p(l_p)$} & The amount of the $l_p$th primary flow allocated on\\
&the link between the $i$th and the $j$th CR routers\\
\hline
$\Upsilon _l$ & The rate of the $l$th secondary flow\\
\hline
$\lambda_{mq}$ & The amount of time share allocated to the $q$th MIS\\
\hline
\end{tabular}
\end{center}
\end{table}

\section{Related Work}
As aforementioned, in literature, the cooperation-based spectrum access is implemented via a cooperative communications based LLC approach where PUs provide SUs with certain spectrum access opportunities in exchange for improved link performance \cite{Simeone2008}. This LLC approach is originally introduced in \cite{Simeone2008} where Simeone et al. demonstrate the feasibility of their proposed scheme via analytical and numerical studies of the Stackelberg games. Later, various cooperative schemes have been proposed for the cooperation-based spectrum access based on this LLC approach. In \cite{Jing2015}, a sequentially observing scheme is proposed to enable PUs to efficiently select SUs as relays in the context of a large number of SUs, and the optimal stopping policy is studied. In \cite{Duan2014}, a contract-based scheme is designed to enable the cooperation-based spectrum access in CRNs. By modeling the PU and the SUs as an employer and employees, Duan et al. study the optimal contract design problem. In \cite{Feng2014}, potential cooperative schemes for a CRN with multiple potentially selfish PUs and SUs are studied. Besides relay SU selection, energy consumption is another factor considered in designing cooperative schemes since it has impact on users' communication strategies. In this view, energy-aware cooperative schemes are investigated based on a sum-constrained power allocation game and a power control game in \cite{Cao2013}. Given limited transmit power of SUs, PUs may need to recruit multiple SUs to relay their data in a multi-hop fashion. In \cite{Li2013}, this multihop relay selection problem is studied based on a network formation game. Motivated by increasing concerns on information security, two types of cooperative schemes are proposed in \cite{Zhang2013} to improve PUs' secrecy rate. As aforementioned, an underlying assumption in the LLC approach is that individual SUs work independently for their own spectrum access opportunities and the generated spectrum access opportunities are exclusively granted to relaying SUs such that other SUs cannot transmit during the cooperation-incurred periods \cite{Long2014,Zhang2014}. Thus, the LLC approach will waste a significant number of spectrum access opportunities to improve the throughput of CRNs. In addition, in the LLC approach, PUs might not actually benefit from SUs' help and thus might not be interested in cooperating with SUs. These observations motivate us to introduce NLC approach in order to enable the cooperation-based spectrum access and boost the capacities of CRNs.

Although the concept of session-based cooperation in cognitive radio networks (CRNs) has been discussed in a few works, such as \cite{Yuan2013} and \cite{Yuan2017}, it is studied from a different perspective from our work. \cite{Yuan2013} and \cite{Yuan2017} primarily address how PUs interact with SUs so that both of them can gain from the cooperation. While our work focus on the interaction between individual SUs. Unlike existing works where SUs independently cooperate with PUs for their own spectrum access opportunities, we advocate the cooperation among SUs based on the observation that SUs can benefits from the collaboration if they are grouped together and collectively cooperate with PUs for the spectrum access opportunities of the group instead of themselves. Different from \cite{Nadkar2011} where non-selfish SUs opportunistically offer their spectrum access opportunities to others with better channel condition, our proposed scheme emphasizes mutual benefits between SUs. That is why we call it a network-level approach and articulate it via the CCHN architecture where CR routers collectively cooperate with PUs for the spectrum access opportunities of the CCHN.

The CCHN architecture is first introduced in our previous work \cite{Pan2012} where the SSP is introduced to provide services for SUs by judiciously deploying CR routers. In such a way, SUs can benefit from cognitive radio technologies even if they do not have CR capability. In \cite{Pan2012}, we demonstrate that the CCHN can efficiently harvest unused licensed bands for service provisioning even in the context of uncertain spectrum availability. Since the performance of our CCHN heavily depends on the placement strategy of CR routers, we design an optimal CR router placement strategy by jointly considering the spectrum and energy efficiency in \cite{Yue2013}. In our recent works, we have verified the effectiveness of the CCHN in supporting spectrum-auction-based spectrum access \cite{Pan2014,Li2016}. Although our previous works have demonstrated the CCHN can efficiently support the spectrum-sensing-based and spectrum-auction-based spectrum access, how to support the cooperation-based spectrum access in the CCHN is still an open problem.


\section{CCHN and Network-Level Session-Based Cooperation}
\subsection{CCHN Architecture}
\begin{figure}[!t]
 \begin{center}
  \includegraphics[width=3in]{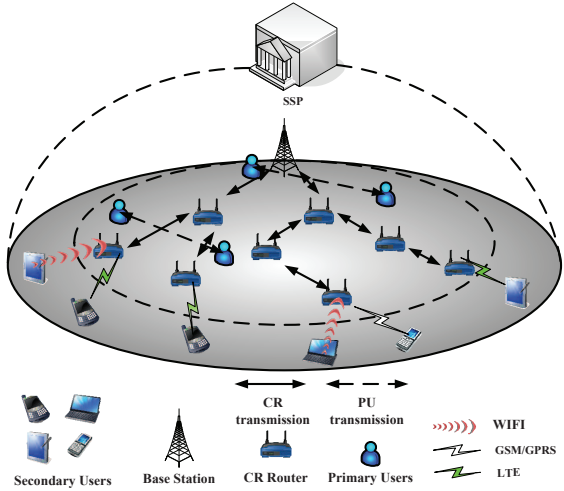}
  \end{center}
  \begin{center}
   \parbox{5cm}{\caption{The CCHN architecture.}  \label{fig2}}
  \end{center}
  \end{figure}

As introduced in our previous work, the CCHN consists of an SSP, BSs, CR routers and SUs as shown in Fig. \ref{fig2} \cite{Pan2012}. The SSP is an independent wireless service provider, such as a cellular operator that is willing to provide better or new types of services to cellular users, and has its own licensed spectrum bands, referred to as the SSP's basic bands. The SSP is in charge of spectrum coordination and service provisioning within its coverage area. To provide communication services to SUs, the SSP deploys or leases some BSs for fundamental service coverage as done in cellular systems and CR routers for efficient resource utilization. BSs are interconnected with wired connections via Internet or other high-speed data networks and work as gateways for CR routers, so that the CCHN can gain backbone network services. BSs also serve as an agent for the SSP to exchange control signaling with CR routers and SUs. CR routers are intelligent wireless routers with cognitive capability and operate under the supervision of the SSP. Both BSs and CR routers are equipped with multiple radio interfaces, such as cognitive radio interface, cellular interface, and WiFi interface, and can operate over the SSP's basic bands, unlicensed bands (e.g., ISM bands), and unoccupied licensed bands. CR routers form a cognitive radio mesh network to help the SSP deliver services to SUs in collaboration with BSs. SUs are wireless terminals or devices (e.g., smart phones and laptops) obtaining services via certain access technologies (e.g., GSM/GPRS, LTE and WiFi) and may or may not have cognitive capability. SUs access the SSP's services by connecting to CR routers or BSs, and CR routers directly connecting to SUs are called edge CR routers. If SUs have cognitive capability, they can communicate with edge CR routers via both their basic access technologies and cognitive radios\footnote{By basic access technologies, we refer to the communication technologies which SUs normally use to get communication services. For example, for cellular users, their basic access technologies can be either GSM or LTE.}. If SUs' devices do not have cognitive radio interfaces, edge CR routers will tune to the interfaces which SUs normally use to deliver services. Each edge CR router constantly collects data requests in its coverage area and submits those collected data requests to the SSP for resource allocation. Based on the data requests and available resources, the SSP carries out network optimization, and the decisions will be delivered to CR routers via the SSP's basic bands\footnote{The SSP will reserve a certain number of basic bands for the control message exchange among BSs and CR routers. Meanwhile, the SSP will allocate a certain number of basic bands to enable SUs to access the CCHN. Then, the remaining basic bands will be allocated to the cognitive radio mesh of CR routers for data delivery.}. Under the guidance of the SSP, BSs and CR routers collectively build up paths to deliver services to SUs via multi-hop transmissions. As shown in \cite{Pan2012,Pan2014}, the CCHN architecture is very flexible in supporting various types of spectrum-sharing paradigms, including spectrum-sensing-based and spectrum-auction-based spectrum sharing. In this paper, we consider the use of the CCHN to support another paradigm, i.e., cooperation-based spectrum sharing.

\subsection{Proposed NLC Scheme}
Under the proposed NLC scheme, when running out of available bands,\footnote{As aforementioned, certain number of basic bands are reserved for control signaling.} the SSP will coordinate BSs and CR routers to cooperate with PUs to gain spectrum access opportunities. The SSP directly obtains lengths and data volumes of primary sessions from the PSP and makes cooperating decisions on different primary sessions via network optimization. Once the SSP decides to cooperate with a primary session, it supervises its BSs and CR routers to build up high-speed paths for this primary session to expedite E2E primary service delivery. After a scheduled primary service is delivered, the remaining time of the primary session during the intended transmission period is granted to the SSP for spectrum access. Then, the SSP intelligently allocates cooperation-incurred spectrum access opportunities among its BSs and CR routers to efficiently serve SUs.

The NLC scheme has a number of appealing features to address all challenges mentioned in Section I. First, it is the SSP instead of SUs who involves in the cooperating process and the complexity of cooperation is shifted from SUs to the network. The SSP directly interacts with the PSP for cooperation-related information exchange and supervises operations of secondary network facilities, i.e., BSs and CR routers, via basic bands, for efficient resource utilization. Second, under the supervision of the SSP, secondary network facilities, as a group, cooperate with PUs for spectrum access opportunities of the CCHN, i.e., the SSP, and the SSP can optimally allocate the cooperation-incurred spectrum access opportunities among secondary network facilities so that the network capacity is significantly improved. Third, in our NLC scheme, the SSP coordinates secondary network facilities to expedite data transmissions in primary sessions in order to gain spectrum access opportunities. As a result, PUs will enjoy better services since the latency of E2E service delivery is significantly improved with the help of the SSP and are more likely to join the cooperation-based spectrum access under our NLC scheme.

\section{Network Model}
In this section, we will introduce the basic network configuration as well as the related communication models. To examine the effectiveness of our NLC scheme, we do not consider the SSP's basic bands in the following analysis.
\subsection{Network Configuration}
We consider a typical CCHN with a BS, denoted as $b$, and $N$ CR routers which are deployed by the SSP. Those CR routers are indexed as $\mathcal{N}=\left\{ {1,2, \cdots ,N} \right\}$ and $L$ of them are edge CR routers denoted as $s\left( l \right)$, $l \in \mathcal{L}$, where $\mathcal{L}=\left\{ {1,\cdots,L} \right\}$, and $s(l) \in \mathcal{N}$. The set of secondary network facilities is ${\mathcal{N}_s} =\mathcal{N} \cup \{b\} $. There are $L_p$ active primary sessions collocated with the CCHN. $s_p \left( l_p \right)$ and $d_p \left( l_p \right)$ represents the source and the destination of the $l_p$th primary session, respectively, where ${l_p} \in {\mathcal{L}_p} = \left\{ {1, \cdots ,{L_p}} \right\}$. Unlike previous works, in this paper, the lengths and data volumes of different primary sessions are allowed to be different. The length and the data volume of the $l_p$th primary session is denoted as $T_{l_p}$ and $D_{l_p}$, respectively. Without loss of generality, the cooperation between PUs and the SSP is conducted on a single band, i.e., all the considered primary sessions operate in the same band. Each BS or CR router only has single cognitive radio\footnote{Since the purpose of this paper is to investigate the feasibility of the NLC scheme, we consider the single-channel single-radio case for simplicity.}. We assume that primary sessions do not interfere with each other due to the coordination of the PSP. BSs and CR routers can access the PUs' band only when their transmission activities do not cause harmful interference to ongoing primary sessions. To characterize the interfering relationship among PUs, BSs and CR-routers, we will introduce related communication models in the next subsection.

\subsection{Communication Models}
\subsubsection{Transmission Range and Interference Range}
Our formulation proceeds with the widely adopted protocol model\footnote{The relation between the protocol model and the physical model has been discussed in \cite{Shi2009}, and it has been shown that the protocol model can be accurately transformed into the physical model if the interference range is properly set.} where signal transmission in the physical layer is characterized by a transmission range and an interference range. For example, CR router $j$ is able to successfully receive signals from CR router $i$ if it falls in the transmission range of CR router $i$ and stays outside the interference range of any other transmitting secondary network facilities and PUs. For simplicity, we assume that network entities of the same type employ the same transmit power $P_t^{\mu}$, $\mu \in \left\{C,b,P\right\}$, where $C$ represents CR routers, $b$ represents the BS, $P$ represents PUs, and the subscript $t$ indicates the power is for transmission. For a transmitter of type $\mu$, the received power at the receiver is
\begin{align}
\label{1}
{P_r} = {P_t^{\mu}}\gamma d^{ - n},
\end{align}
where $\gamma$ is the antenna related constant, $n$ is the path loss exponent, and $d$ is the distance between the transmitter and the receiver. The received signal can be correctly decoded at the receiver only when $P_r$ is greater than a predetermined threshold $P_{R}^{\nu}$, where $\nu \in \left\{C,b,P\right\}$ signifies the type of the receiver. Then, the distance between the transmitter and the receiver should satisfy ${P_t^{\mu}}\gamma d^{ - n} \ge {P_{R}^{\nu}}$, which implies the transmission range of a network entity of type $\mu$ to another network entity of type $\nu$ is $R_{T}^{\mu \nu}={\left( {{{\gamma {P_t^{\mu}}} \mathord{\left/
 {\vphantom {{\gamma {P_t^{\mu}}} {{P_{R}^{\nu}}}}} \right.
 \kern-\nulldelimiterspace} {{P_{R}^{\nu}}}}} \right)^{{1 \mathord{\left/
 {\vphantom {1 n}} \right.
 \kern-\nulldelimiterspace} n}}}$. Similar to \cite{Pan2012,Pan2014}, the received interference power is not negligible if it exceeds a threshold $P_I^{\nu}$, $\nu \in \left\{C,b,P\right\}$. Thus, the interference range of a network entity of type $\mu$ to another network entity of type $\nu$ is $R_I^{\mu \nu}={\left( {{{\gamma {P_t^{\mu}}} \mathord{\left/
 {\vphantom {{\gamma {P_t^{\mu}}} {{P_I^{\nu}}}}} \right.
 \kern-\nulldelimiterspace} {{P_I^{\nu}}}}} \right)^{{1 \mathord{\left/
 {\vphantom {1 n}} \right.
 \kern-\nulldelimiterspace} n}}}$.

\subsubsection{Achievable Data Rate}
If CR router $j$ is in the transmission range of CR router $i$, there exists a communication link, denoted as $(i,j)$, between these two routers. The achievable data rate of link $(i,j)$ is a given parameter denoted as $c_{ij}$. Generally, $c_{ij}$ is determined by the channel bandwidth and physical layer techniques, such as multi-antenna techniques, adaptive coding and modulation techniques. Once the physical layer techniques are given, $c_{ij}$ is determined accordingly and used as a constant in the analysis.

\section{Session-Based Cooperative Mechanism Design}
In this section, we will explore the design of cooperative mechanism by jointly considering two tightly coupled problems, primary session selection and efficient resource utilization. On the one hand, whether the SSP chooses to cooperate with a primary session depends not only on the length and data volume of this primary session, but also on how the SSP utilizes the cooperation-incurred transmission opportunity. Clearly, the SSP will gain nothing from cooperating with a primary session whose neighbouring CR routers are never scheduled. On the other hand, how the SSP utilizes the cooperation-incurred periods is affected by primary session selection as well since BSs and CR routers cannot access the band if they interfere with an on-going primary session. Thus, these two problems should be jointly considered when designing the cooperative mechanism. In this view, we study the cooperative mechanism design by jointly considering primary session selection, flow routing and link scheduling to maximize the aggregated throughput of the CCHN. To ease the problem formulation, we use $\theta_{l_p}$ to denote the cooperating decision of the SSP on the $l_p$th primary session, i.e.,
\begin{align}
\label{3}
{\theta _{{l_p}}} = \left\{ {\begin{array}{*{20}{c}{c}}
   1, & \text{the $l_p$th primary session is selected} \\
   0, & \text{the $l_p$th primary session is not selected} \\
\end{array}} \right.
\end{align}

In essence, the cooperative mechanism design considered in this paper is a throughput maximization problem for multi-hop wireless networks with interference constraints\cite{Jain2003}. As proved in Appendix A, the considered problem can be decomposed into a maximal independent set (MIS) searching subproblem and a mix integer linear programming (MILP). The MIS searching subproblem tries to identify a set of MISs of the PU related conflict graph which characterizes the interfering relations among CR links, PU-related links, and primary sessions. Based on the obtained set of MISs, we can formulate the MILP to facilitate final solution finding. In the following, we will develop the considered optimization problem based on this decomposable structure which renders us good heuristics for solution finding.


In the rest of this section, we first introduce the PU-related conflict graph to characterize the interfering relations among CR links, PU-related links and primary sessions. Afterwards, on the basis of the PU-related conflict graph, we mathematically establish link scheduling and flow routing constraints and cast the cooperative mechanism design into a throughput maximization problem with primary session selection. To solve the optimization problem, we propose an augmented SIO-based algorithm to search for MISs.

\subsection{PU-Related Conflict Graph and MISs}
Since flow routing and link scheduling decisions of the SSP are affected by primary session selection, unlike \cite{Pan2012,Pan2014}, our PU-related conflict graph $G=(V,E)$ characterizes interfering relationship not only among different CR links but also among CR links, PU-related links, and primary sessions, where $V$ is the vertex set and $E$ is the edge set. Since PUs will not relay traffic for the SSP, PU-related links only refer to links from sources of primary sessions to CR routers or BSs and links from CR routers or BSs to destinations of primary sessions. Each vertex in the PU-related conflict graph corresponds to a CR link, a PU-related link or a primary session which is represented as an ordered pair. For example, $(i,j)$, $i,j \in \mathcal{N}$ represents the CR link from CR router $i$ to CR router $j$ and exists only when CR router $j$ is within CR router $i$'s transmission range. $\left(s_p\left(l_p\right),j\right)$, $l_p \in \mathcal{L}_p, j \in \mathcal{N}$, signifies the PU-related link from the source of the $l_p$th primary session to CR router $j$. $\left(s_p\left(l_p\right),d_p\left(l_p\right)\right)$, $l_p \in \mathcal{L}_p$ represents the $l_p$th primary session.

Similar to \cite{Li2010,Pan2012,Pan2014}, two communication links are said to be conflicting if the receiver of a link is within the interference range of the transmitter of the other link. This condition covers the following three kinds of conflicting relationship between different links:
\begin{enumerate}
  \item Two links sharing the same transmitter or receiver.
  \item The receiver of a link is the transmitter of another link.
  \item Two links do not share a radio, but the transmission of a link will interfere with the reception of the other link.
\end{enumerate}
The first conflicting relationship implies a single radio cannot support multiple concurrent transmissions/receptions on the same band. The second one means a single radio cannot use the same band for simultaneous transmission and reception. The last one is due to co-channel interference. Based on the conflicting condition for communication links, a CR link or a PU-related link conflicts with a primary session once it conflicts with any primary link, i.e., links between PUs, used by this primary session. According to those defined conflicting relationship, we add an undirected edge between two vertices in $V$ if their corresponding links conflict with each other.

 \begin{figure}[!t]
 \begin{center}
 \subfigure[The toy topology of a CCHN.]{
\begin{minipage}[b]{0.4\textwidth}
\includegraphics[width=1\textwidth]{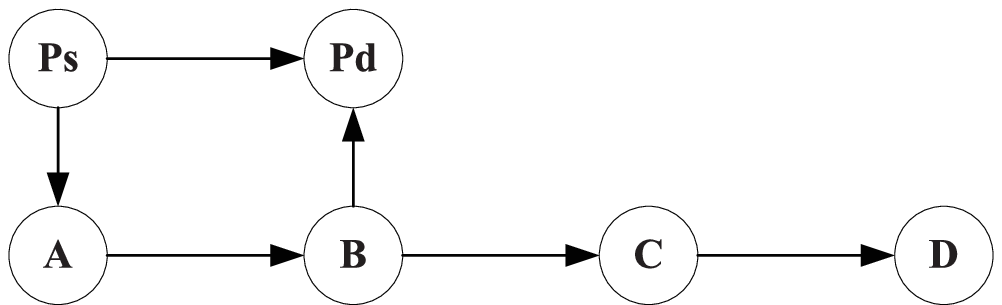}
\end{minipage}
}
\subfigure[PU-related conflict graph.]{
\begin{minipage}[b]{0.35\textwidth}
\includegraphics[width=1\textwidth]{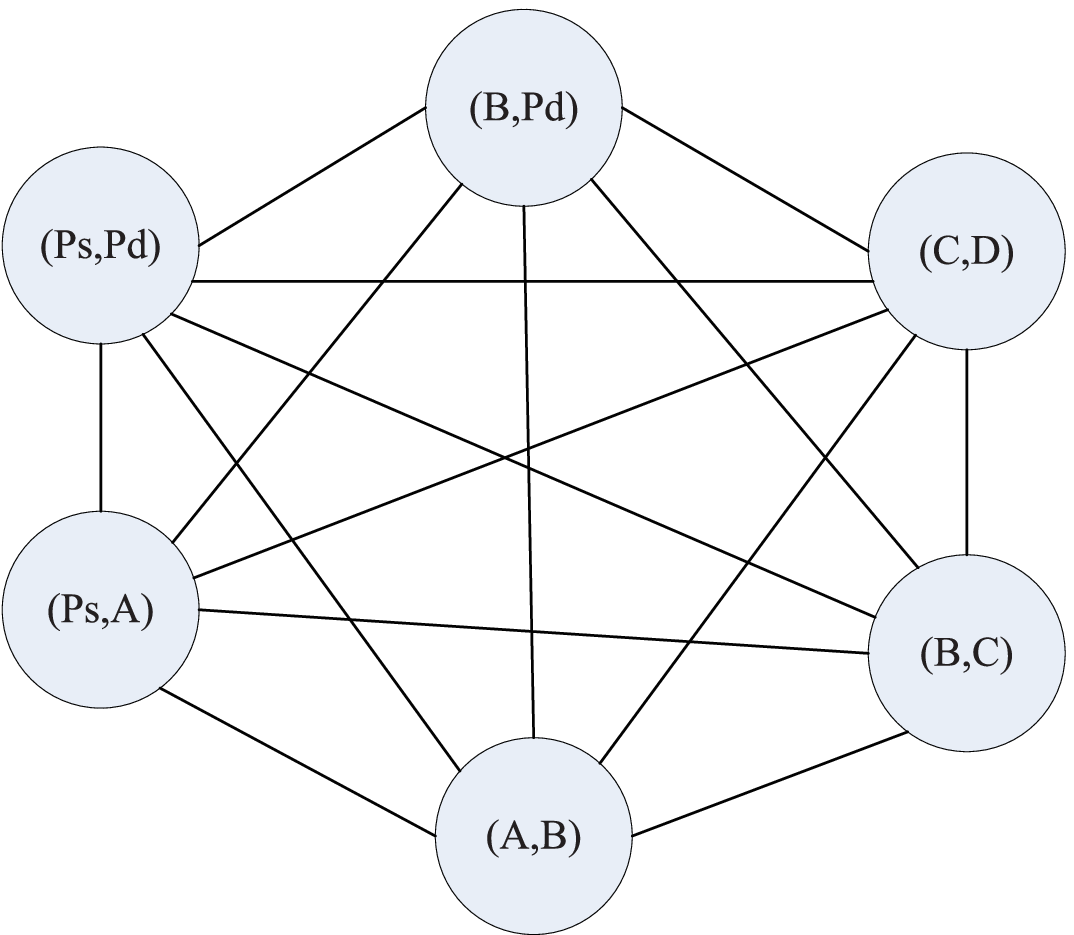}
\end{minipage}
}
  \end{center}
  \begin{center}
   \parbox{7cm}{\caption{PU-related conflict graph for a toy CCHN.}  \label{fig3}}
  \end{center}
  \end{figure}

For illustrative purpose, we use a toy CCHN shown in Fig. \ref{fig3} to show how to construct a PU-related conflict graph. The toy CCHN consists of four CR routers, i.e., $A$, $B$, $C$ and $D$, and coexists with a primary session with $\rm{Ps}$ as the source and $\rm{Pd}$ as the destination. For convenience, as commonly done in the literature, we assume $R_T^{\mu \nu}=R_T$, $\forall \mu,\nu \in \left\{C,b,P\right\}$, $R_I^{\mu \nu}=R_I$, $\forall \mu,\nu \in \left\{C,b,P\right\}$, and $d({\rm{Ps}},{\rm{Pd}})=d({\rm{Ps}},A)=d(B,{\rm{Pd}})=d(A,B)=d(B,C)=d(C,D)=R_T=0.5R_I$, where $d(A,B)$ is the Euclidian distance between $A$ and $B$. Under this assumption, $\rm{Ps}$ can directly reach $\rm{Pd}$, and there are $6$ CR links and $2$ PU-related links in the toy CCHN\footnote{The primary session is assumed to be implemented via single-hop transmissions for simplicity. Our formulation can incorporate both primary sessions implemented via single-hop transmissions and those implemented via multi-hop transmissions. Particularly, in performance evaluation part, we have considered a scenario where some primary sessions are implemented via multi-hop transmissions.}. For simplicity, we construct the PU-related conflict graph based on the links indicated in Fig. \ref{fig3}(a) and the result is shown in Fig. \ref{fig3}(b). There exists an edge between vertices $(\rm{Ps},\rm{Pd})$ and $(C,D)$ since $\rm{Pd}$ is within the interference range of $C$. There is an edge between vertices $(B,\rm{Pd})$ and $(B,C)$ because they share the same transmitter $B$. There is an edge between vertices $(A,B)$ and $(B,C)$ as $B$ is the receiver in $(A,B)$ and the transmitter in $(B,C)$. The same arguments apply to other vertices as well.

Given a set of vertices $I \subseteq V$, if any two vertices in $I$ do not share an edge, the corresponding links and primary sessions in $I$ can be scheduled simultaneously without interfering with each other. In this case, this set of vertices is called an independent set. If adding one more vertex into the independent set $I$ results in a non-independent set, the set $I$ is called the maximal independent set (MIS). By scheduling corresponding links of an MIS, we can accommodate as many communication links as possible, which improves frequency reuse. We collect all MISs of the PU-related conflict graph in a set $\mathcal{I}=\{I_1,\cdots,I_q,\cdots,I_Q\}$, where $I_q$ is the $q$th MIS, $Q$ is the total number of MISs and equals to the cardinality of $\mathcal{I}$, i.e., $|\mathcal{I}|$. Based on the $l_p$th primary session, we divide $\mathcal{I}$ into $\mathcal{I}_{l_p}$ and $\overline{\mathcal{I}}_{l_p}$ with ${\mathcal{I}_{{l_p}}}  \cap \overline{\mathcal{I}}_{{l_p}}=\emptyset$ and ${\mathcal{I}_{{l_p}}}  \cup \overline{\mathcal{I}}_{{l_p}}=\mathcal{I}$, where $\mathcal{I}_{l_p}$ is the set of MISs which include the vertex corresponding to the $l_p$th primary session. In what follows, we will formulate our throughput maximization problem based on MISs of the PU-related conflict graph.

\subsection{Flow Routing and Link Scheduling Constraints}
To optimally utilize network resources, we should jointly consider flow routing and link scheduling which are tightly coupled problems. On the one hand, the scheduling at the data link layer should be able to support the flows at the network layer. On the other hand, how much flow can be carried at the network layer is determined by the scheduling at the data link layer. In this subsection, we mathematically formulate the flow routing and link scheduling constraints for the NLC scheme. To embrace possible cooperation between primary sessions and the SSP, unlike existing works, we add extra constraints for PU-related links and incorporate primary session selection and variations in lengths of primary sessions into our formulation.

\subsubsection{Control Interval}
Usually, the SSP makes link scheduling and routing decisions during a certain period of time which is called the control interval. For the sake of efficient resource utilization, the SSP should choose its control interval based on the lengths of primary sessions. Unlike its counterpart at physical/MAC layer, determining control interval is more involved in the NLC scheme since different primary sessions may have different lengths and it is unknown if the SSP cooperates with a primary session.

\begin{figure}[!t]
 \begin{center}
  \includegraphics[width=3in]{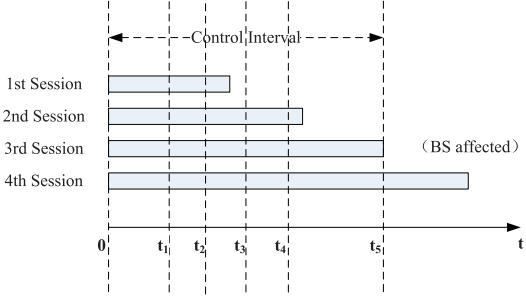}
  \end{center}
  \begin{center}
   \parbox{8cm}{\caption{The selection of the control interval. ($t_m$ represents the promised finishing time for the $m$th primary session if the SSP cooperates, defined later)}  \label{fig4}}
  \end{center}
  \end{figure}

In the CCHN, SUs' data are delivered to the BS where the connections to data networks are provided. As a result, to exploit the cooperation-incurred spectrum access opportunities for service delivery, the SSP must cooperate with the primary sessions whose activities will conflict with that of the BS so that SUs' data can be delivered to the BS during the cooperation-incurred periods. Suppose multiple primary sessions exist in the vicinity of the BS and let $T_{min}$ denote the lengths of the shortest primary sessions conflicting with the activity of the BS. Considering the uncertainty in PUs' activities, it makes no sense for the SSP to make scheduling for a period longer than $T_{min}$ as it may not be able to access data networks afterwards. Thus, it is reasonable to set the length of the control interval as $T=T_{min}$. For clarity, let us consider an example shown in Fig. \ref{fig4}, where $4$ primary sessions exist and the activity of the BS happens to be affected by the $3$rd primary session. In this case, we set the length of the control interval as the length of the $3$rd primary session. In the following formulation, we will regard $T$ as a given parameter.

\subsubsection{Flow Routing Constraints}
To study how much the SSP can gain from cooperating with primary sessions, we consider multi-path routing in this paper. According to our NLC scheme, the SSP should first help PUs finish their transmissions and then utilize the cooperation-incurred periods to serve SUs. When reflected at the network layer, there are two kinds of data flows to be carried over the CCHN. The flows originated from the edge CR routers are referred to as secondary flows, and the flows generated by the primary sessions are called primary flows. For secondary flows, the achievable rate depends on what the network can provide since secondary data traffic is transmitted during cooperation-incurred periods. To encourage PUs to join the cooperating process, the SSP must ensure PUs' data is delivered before a certain time, which implies certain flow rates should be assured for primary flows. Given different rate requirements, the flow routing constraints for secondary and primary flows are introduced separately.

Let $\Upsilon_l$ be the rate of the $l$th secondary flow which is originated from the $l$th edge CR router $s(l)$. We have the following flow conservation constraints at the source $s(l)$ as
\begin{align}
\label{10}
\sum\limits_{j \in {\mathcal{T}_{s\left( l \right)}}} {{f_{s\left( l \right)j}}\left( l \right)}  = {\Upsilon _l}, \\
\label{11}
\sum\limits_{j \in {\mathcal{R}_{s\left( l \right)}}} {{f_{js\left( l \right)}}\left( l \right)}  = 0,
\end{align}
where ${f_{ij}}\left( l \right)$ is the rate of the $l$th secondary flow over link $(i,j)$ ($l \in \mathcal{L}$, $i,j \in \mathcal{N}_s$). (\ref{10}) implies the rate of the flow originated from $s(l)$ is limited by what the network can support. (\ref{11}) guarantees no flow comes back to the source. ${{\mathcal{T}_{s\left( l \right)}}}$ is the set of secondary network facilities within $s(l)$'s transmission range. ${\mathcal{R}_{s\left( l \right)}}$ is the set of secondary network facilities with $s(l)$ in their transmission ranges, i.e., ${\mathcal{R}_{s\left( l \right)}}=\left\{ {j \in {\mathcal{N}_s}\left| {{s(l)} \in {\mathcal{T}_j}} \right.} \right\}$.

If CR router $i$ is an intermediate relay of the $l$th secondary flow, i.e., $i \in \mathcal{N}_s$, $i \ne s(l)$ and $i \ne b$, the flow into $i$ must equal to the flow out of $i$. That is,
\begin{align}
\label{12}
\sum\limits_{j \in {\mathcal{T}_i}} {{f_{ij}}\left( l \right)}  = \sum\limits_{j \in {\mathcal{R}_i}} {{f_{ji}}\left( l \right)}.
\end{align}

In the CCHN, all secondary flows go through the BS for Internet services, which implies the BS $b$ is the common destination for secondary flows. For the $l$th secondary flow, the constraints at $b$ can be formulated as
\begin{align}
\label{13}
\sum\limits_{j \in {\mathcal{R}_b}} {{f_{jb}}\left( l \right)}  = {\Upsilon _l} \\
\label{14}
\sum\limits_{j \in {\mathcal{T}_b}} {{f_{bj}}\left( l \right)}  = 0.
\end{align}

By adding (\ref{12}) for all intermediate relays, we have
\begin{align}
\label{15}
\sum\limits_{j \in {\mathcal{T}_{s\left( l \right)}}} {{f_{s\left( l \right)j}}\left( l \right)}  + \sum\limits_{j \in {\mathcal{T}_b}} {{f_{bj}}\left( l \right)}  = \sum\limits_{j \in {\mathcal{R}_{s\left( l \right)}}} {{f_{js\left( l \right)}}\left( l \right)}  + \sum\limits_{j \in {\mathcal{R}_b}} {{f_{jb}}\left( l \right)}.
\end{align}
From (\ref{15}), if (\ref{10}), (\ref{11}) and (\ref{14}) are given, (\ref{13}) must be satisfied. Therefore, it is sufficient to adopt (\ref{10}), (\ref{11}) and (\ref{14}) in the flow routing constraints for secondary flows.

Unlike traditional network flow problems, PU-related links will not carry any secondary flows. Thus, the $l$th ($l \in \mathcal{L}$) secondary flow over PU-related links must be $0$, i.e., ${f_{{s_p}\left( {{l_p}} \right)j}}\left( l \right) = {f_{i{d_p}\left( {{l_p}} \right)}}\left( l \right) = 0$, $j \in {\mathcal{T}_{s_p\left( l_p \right)}}, i \in {\mathcal{R}_{d_p\left( l_p \right)}}, l_p \in \mathcal{L}_p$.

Besides above constraints, the NLC scheme requires the SSP help PUs finish their transmissions before utilizing the band for its own data transmissions, which implies that certain flow rate should be guaranteed for primary flows. Consequently, for the $l_p$th primary flow, which is generated by the $l_p$th primary session, the constraint at the source ${s_p}\left( {{l_p}} \right)$ can be written as
\begin{align}
\label{16}
\sum\limits_{j \in {\mathcal{T}_{{s_p}\left( {{l_p}} \right)}}} {{f_{{s_p}\left( {{l_p}} \right)j}^p}\left( l_p \right)}  \ge \frac{{{\theta _{{l_p}}}{D_{{l_p}}}}}{T},
\end{align}
where ${f_{s_p\left({l_p}\right)j}^p}\left( {{l_p}} \right)$ is the rate of the $l_p$th primary flow allocated over the link from the source of the $l_p$th primary session to CR router $j$ ($j \in \mathcal{N}_s$), $\mathcal{T}_{{s_p}\left( {{l_p}} \right)}$ is the set of secondary network facilities within the transmission range of ${s_p}\left( {{l_p}} \right)$, ${{\theta _{{l_p}}}}$ is a $0-1$ parameter representing the SSP's decision on whether to cooperate with the $l_p$th primary session. $D_{l_p}$ is the data volume of the $l_p$th primary session. $T$ is the length of the control interval. Since we have already precluded the links from secondary network facilities to ${s_p}\left( {{l_p}} \right)$ during the construction of the conflict graph, it is not necessary to include another constraint similar to (\ref{11}) for primary flows.

Similar to (\ref{12}), if CR router $i$ is an intermediate relay of the $l_p$th primary flow, then,
\begin{align}
\label{17}
\sum\limits_{j \in {\mathcal{T}_i} \cup {\omega_{i,T}}} {{f_{ij}^p}\left( {{l_p}} \right)}  = \sum\limits_{j \in {\mathcal{R}_i} \cup {\omega_{i,R}}} {{f_{ji}^p}\left( {{l_p}} \right)},
\end{align}
where ${{\omega_{i,T}}}=\emptyset$ if $d_p\left(l_p \right)$ is outside the transmission range of CR router $i$ and ${{\omega_{i,T}}}=\left\{d_p\left(l_p \right) \right\}$, otherwise. Likewise, ${{\omega_{i,R}}}=\emptyset$ if CR router $i$ is out of the transmission range of $s_p\left(l_p \right)$ and ${{\omega_{i,R}}}=\left\{s_p\left(l_p \right) \right\}$, otherwise. ${f_{ij}^p}\left( l_p \right)$ is the rate of the $l_p$th primary flow over CR link $(i,j)$ ($l_p \in \mathcal{L}_p$, $i,j \in \mathcal{N}_s$), and ${f_{id_p\left({l_p}\right)}^p}\left( {{l_p}} \right)$ represents the flow rate of the $l_p$th primary session over the link from CR router $i$ ($i \in \mathcal{N}_s$) to the destination of the $l_p$th primary session.

Like the secondary flow case, the constraint at the destination $d_p\left(l_p \right)$ will be automatically satisfied once (\ref{16}) and (\ref{17}) hold. As a result, this constraint is not listed.

Noticing that PU-related links $({{s_p}\left( {{l_p}} \right),j})$'s and $({i,{d_p}\left( {{l_p}} \right)})$'s, $j \in {\mathcal{T}_{s_p\left( l_p \right)}}, i \in {\mathcal{R}_{d_p\left( l_p \right)}}, l_p \in \mathcal{L}_p$ will not relay traffic for other primary flows, we require ${f_{{s_p}\left( {{l_p}} \right)j}^p}\left( {{l_p}'} \right) = {f_{i{d_p}\left( {{l_p}} \right)}^p}\left( {{l_p}'} \right) = 0$ when ${l_p} \ne {l_p}'$ and ${l_p}, {l_p}' \in \mathcal{L}_p$.
\subsubsection{Link Scheduling Constraints}
In this paper, we consider time based link scheduling where different links are allocated with certain periods of time to build up flows between end systems. Consequently, the flow rates which the network layer can provide depend on the data rate of each link as well as the time share allocated to these links. To provide PUs with incentives to cooperate, the SSP must guarantee PUs' data is delivered to the destination earlier than what would have been scheduled without the SSP's help. We introduce an incentive parameter $\alpha$ to capture this point and assume the SSP will consider cooperating with the $l_p$th primary session only if it can at least deliver the data of the $l_p$th primary session to the destination during a period of ${{{T_{{l_p}}}} \mathord{\left/
 {\vphantom {{{T_{{l_p}}}} \alpha }} \right.
 \kern-\nulldelimiterspace} \alpha }$. In practice, $\alpha$ can either be determined by the PSP who proactively looks for cooperation or be set by the SSP who wishes to cooperate with PUs for spectrum access opportunities. As mentioned above, if the SSP decides to cooperate with the $l_p$th primary session, the CCHN has to support a network flow with rate ${{{D_{{l_p}}}} \mathord{\left/
 {\vphantom {{{D_{{l_p}}}} T}} \right.
 \kern-\nulldelimiterspace} T}$. In this case, the incentive mechanism demands a link scheduling which is able to build up a flow with rate ${{{D_{{l_p}}}} \mathord{\left/
 {\vphantom {{{D_{{l_p}}}} T}} \right.
 \kern-\nulldelimiterspace} T}$ for the $l_p$th primary session during a period of ${{{T_{{l_p}}}} \mathord{\left/
 {\vphantom {{{T_{{l_p}}}} \alpha }} \right.
 \kern-\nulldelimiterspace} \alpha }$.

Without loss of generality, we assume primary sessions are sorted by session times and the $l_p$th primary session has the $l_p$th shortest duration, i.e., ${T_1} \le {T_2} \le  \cdots  \le {T_{{L_p}}}$. Our formulation needs another set of parameters defined as\footnote{We assume all primary sessions start at $0$ for simplicity. In practice, the start time can be determined by the SSP.}
\begin{align}
\label{34}
{t_m} = \left\{ {\begin{array}{*{20}{c}}
   {\begin{array}{*{20}{c}}
   0  \\
   {\min \left\{ {{{{T_m}} \mathord{\left/
 {\vphantom {{{T_m}} \alpha }} \right.
 \kern-\nulldelimiterspace} \alpha },T} \right\}}  \\
   T  \\
\end{array}} & {\begin{array}{*{20}{c}}
   {m = 0}  \\
   {1 \le m \le {L_p}}  \\
   {m = {L_p} + 1}  \\
\end{array}}  \\
\end{array}} \right.,
\end{align}
where $t_m$ ($m=1,\cdots,L_p$) corresponds to the finishing time of the $m$th primary session promised by the SSP. If ${{{{T_m}} \mathord{\left/
 {\vphantom {{{T_m}} \alpha }} \right.
 \kern-\nulldelimiterspace} \alpha }} \ge T$, the incentives for PUs will be automatically satisfied when the SSP delivers their data during the control interval of length $T$. In view of that, it is enough to set $t_m$ as ${\min \left\{ {{{{T_m}} \mathord{\left/
 {\vphantom {{{T_m}} \alpha }} \right.
 \kern-\nulldelimiterspace} \alpha },T} \right\}}$ for $m=1,\cdots,L_p$. To facilitate the mathematical formulation of the incentive-related constraints, we divide the control interval based on $t_m$'s defined in (\ref{34}). For example, in Fig. \ref{fig4}, the control interval is divided by $\left\{ {{t_1},{t_2},{t_3},{t_4},{t_5}} \right\}$, where ${t_1}$, ${t_2}$, ${t_3}$, ${t_4}$ are the promised finishing time of the four primary sessions and $t_5=T$ is the end of the control interval. Since the data of the $m$th primary session must be delivered before $t_m$, different set of flows are carried by the CCHN during the intervals $\left( {t_{m-1},{t_m}} \right)$, $m=1,\cdots,L_p+1$. In this view, we establish separate link scheduling constraints for these intervals.

As mentioned before, at any time, at most one MIS in $\mathcal{I}$ can be scheduled to transmit. To proceed, define $0 \le {\lambda _{mq}} \le 1$ as the time share allocated to the $q$th MIS $I_q$ in the interval $\left( {{t_{m - 1}},{t_m}} \right)$, $m=1,\cdots,L_p+1$. Then, we have our first set of link scheduling constraints
\begin{align}
\label{5}
\sum\limits_{q = 1}^Q {{\lambda _{mq}}}  \le \frac{{{t_m} - {t_{m - 1}}}}{T},m = 1, \cdots ,{L_p} + 1.
\end{align}

To protect primary sessions, the links conflicting with the $l_p$th primary session can be scheduled only when the SSP chooses to cooperate with the $l_p$th primary session. That is, the SSP can schedule the MISs in $\overline{\mathcal{I}}_{{l_p}}$ if it decides to cooperate with the $l_p$th primary session. Otherwise, only the MISs in $\mathcal{I}_{l_p}$ can be scheduled. Consequently, we have the following constraint related to the $l_p$th ($l_p \in \mathcal{L}_p$) primary session in interval $\left( {t_{m-1},{t_m}} \right)$, $m=1,\cdots,L_p+1$
\begin{align}
\label{6}
\sum\limits_{q = 1}^Q {{\lambda _{mq}}} 1\left( {{I_q} \in {{\overline {\mathcal{I}} }_{{l_p}}}} \right) \le & 1\left( {{T_{{l_p}}} \ge {t_{m-1}}} \right){\theta _{{l_p}}} \nonumber \\ &\times \frac{{\min \left\{ {{T_{{l_p}}} - {t_{m - 1}},{t_m} - {t_{m - 1}}} \right\}}}{T},
\end{align}
where ${1\left( {{I_q} \in \overline{\mathcal{I}}_{{l_p}}} \right)}=1$ if ${I_q}$ belongs to $\overline{\mathcal{I}}_{{l_p}}$, otherwise, ${1\left( {{I_q} \in \overline{\mathcal{I}}_{{l_p}}} \right)}=0$. $1\left( {{T_{{l_p}}} \ge {t_{m-1}}} \right)$ is an indicator function which signifies the MISs in $\overline{\mathcal{I}}_{{l_p}}$ cannot be scheduled after $T_{l_p}$, the duration of the intended transmission periods of the $l_p$th session. When the SSP decides to cooperate with the $l_p$th primary session, ${{\theta _{{l_p}}}}=1$, and at most $\frac{{\min \left\{ {{T_{{l_p}}} - {t_{m - 1}},{t_m} - {t_{m - 1}}} \right\}}}{T}$ can be assigned to the MISs in $\overline{\mathcal{I}}_{{l_p}}$. The min operation in (\ref{6}) is used to cover the case where $t_{m-1}<T_{l_p}<t_m$ (e.g., $t_2<T_1<t_3$ in Fig. \ref{fig4}). If the SSP chooses not to cooperate with the $l_p$th primary session, ${{\theta _{{l_p}}}}=0$. Since $\lambda_{mq} \ge 0$, (\ref{6}) forces $\lambda_{mq} = 0, \forall  I_q \in \overline{\mathcal{I}}_{{l_p}}$, i.e., MISs in $\overline{\mathcal{I}}_{{l_p}}$ cannot be scheduled if the SSP does not cooperate with the $l_p$th primary session.

Since a flow is feasible only when there exists a schedule of the links to support it, we need a few more constraints to relate flow rate to link scheduling. To mathematically formulate these constraints, we denote the data rate for CR link $(i,j)$, $i,j \in \mathcal{N}_s$, when $I_q$ is scheduled as $r_{ij}\left(I_q\right)$ which is defined as
\begin{align}
\label{7}
{r_{ij}}\left( {{I_q}} \right) = \left\{ {\begin{array}{*{20}{c}{c}}
   {{c_{ij}}} & (i,j) \in I_q \\
   0 & (i,j) \notin I_q \\
\end{array}} \right..
\end{align}
$c_{ij}$ is the achievable data rate for CR link $(i,j)$. If $I_q$ is assigned $\lambda_{mq}$ of the whole control interval during $\left(t_{m-1},t_m\right)$, the flow rate contributed by scheduling $I_q$ over $\left(t_{m-1},t_m\right)$ is $\lambda_{mq} r_{ij}\left(I_q\right)$. Following (\ref{7}), we can define similar parameters for PU-related links.

As mentioned above, the data of the $l_p$th primary session must be delivered before $t_{l_p}$, $l_p \in \mathcal{L}_p$, which implies the SSP must be able to build up the $l_p$th primary flow in the CCHN merely based on the link scheduling in $\left( {0,{t_{{l_p}}}} \right)$. Together with the fact that the data of the $k$th ($1 \le k \le l_p$) primary session has already been delivered at the time $t_{l_p}$, we have the following set of constraints for CR link $(i,j)$, $i,j \in \mathcal{N}_s$
\begin{align}
\label{35}
\sum\limits_{k = 1}^{{l_p}} {f_{ij}^p\left( k \right)}  \le \sum\limits_{m = 1}^{{l_p}} {\sum\limits_{q = 1}^Q {{\lambda _{mq}}{r_{ij}}\left( {{I_q}} \right)} }, l_p \in \mathcal{L}_p.
\end{align}
For example, when $l_p=1$, (\ref{35}) reduces to $f_{ij}^p\left( 1 \right) \le \sum\limits_{q = 1}^Q {{\lambda _{1q}}{r_{ij}}\left( {{I_q}} \right)}$ which means the data of the $1$st primary session has been delivered by $t_1$.

Unlike primary flows, secondary flows are carried by the leftover network resources. Thus, their rates depend on both the rates of primary flows and the amount of network resources which can be provided by the CCHN during the whole control interval. This provides us with the following constraints for CR link $(i,j)$, $i,j \in \mathcal{N}_s$
\begin{align}
\label{36}
\sum\limits_{{l_p} = 1}^{{L_p}} {f_{ij}^p\left( {{l_p}} \right)}  + \sum\limits_{l = 1}^L {{f_{ij}}\left( l \right)}  \le \sum\limits_{m = 1}^{{L_p} + 1} {\sum\limits_{q = 1}^Q {{\lambda _{mq}}{r_{ij}}\left( {{I_q}} \right)} },
\end{align}
where the left side is the sum of flow rates to be supported by the network and the right-hand side presents what the link scheduling can provide.

For PU-related links, we can easily write down the same set of constraints as (\ref{35}) and (\ref{36}). As stated in the flow routing part, each PU-related link will not relay traffic for either the SSP or other primary sessions. As a result, we can simplify the constraints for PU-related links as ($l_p \in \mathcal{L}_p$)
\begin{align}
\label{9}
 &f_{{s_p}\left( {{l_p}} \right)j}^p\left( {{l_p}} \right) \le \sum\limits_{m = 1}^{{l_p}} {\sum\limits_{q = 1}^Q {{\lambda _{mq}}{r_{{s_p}\left( {{l_p}} \right)j}}\left( {{I_q}} \right)} }  \\
\label{37}
 &f_{i{d_p}\left( {{l_p}} \right)}^p\left( {{l_p}} \right) \le \sum\limits_{m = 1}^{{l_p}} {\sum\limits_{q = 1}^Q {{\lambda _{mq}}{r_{i{d_p}\left( {{l_p}} \right)}}\left( {{I_q}} \right)} }  \\
\label{38}
 &f_{{s_p}\left( {{l_p}} \right)j}^p\left( {{l_p}} \right) \le \sum\limits_{m = 1}^{{L_p} + 1} {\sum\limits_{q = 1}^Q {{\lambda _{mq}}{r_{{s_p}\left( {{l_p}} \right)j}}\left( {{I_q}} \right)} }  \\
\label{39}
 &f_{i{d_p}\left( {{l_p}} \right)}^p\left( {{l_p}} \right) \le \sum\limits_{m = 1}^{{L_p} + 1} {\sum\limits_{q = 1}^Q {{\lambda _{mq}}{r_{i{d_p}\left( {{l_p}} \right)}}\left( {{I_q}} \right)} },
\end{align}
where (\ref{9}) and (\ref{37}) correspond to (\ref{35}), and (\ref{38}) and (\ref{39}) correspond to (\ref{36}). Since (\ref{38}) and (\ref{39}) will be automatically satisfied once (\ref{9}) and (\ref{37}) are valid, only (\ref{9}) and (\ref{37}) are listed as constraints for our throughput maximization problem.

\subsection{Cooperative Mechanism Design under Multiple Constraints}
In our NLC scheme, to serve SUs, the SSP uses its own network resources to cooperate with primary sessions to obtain spectrum access opportunities. To exploit the cooperation-incurred benefits, the SSP seeks optimal strategies to select appropriate primary sessions to cooperate with, choose secondary network facilities to relay PUs' data, assign cooperation-incurred periods to BSs and CR routers for data transmissions, and route secondary flows such that the total throughput of the CCHN is maximized. With the flow routing and link scheduling constraints introduced in Section V.B, the cooperative mechanism design can be cast into the following throughput maximization problem under multiple constraints as
\begin{align}
\text{maximize}\sum\limits_{l \in \mathcal{L}} {{\Upsilon _l}} \nonumber
\end{align}
\begin{align}
\text{s.t.:}\text{ }\text{ }\text{ } & (\ref{10}) \sim (\ref{12}), (\ref{14}), (\ref{16}), (\ref{17}) \nonumber  \\
&(\ref{5}), (\ref{6}), (\ref{35}) \sim (\ref{37})  \nonumber \\
\label{29}
&{f_{ij}}\left( l \right) \ge 0 \text{ }\text{ } \left(l \in \mathcal{L}, i \in \mathcal{N}_s, j \in \mathcal{T}_i\right) \\
\label{30}
&{f_{ij}^{p}}\left( l_p \right) \ge 0 \text{ }\text{ } \left(l_p \in \mathcal{L}_p, i \in \mathcal{N}_s, j \in \mathcal{T}_i\right) \\
\label{31}
&{f_{{{s_p}\left( {{l_p}} \right)}j}^p}\left( l_p \right) \ge 0 \text{ }\text{ } \left(l_p \in \mathcal{L}_p, j \in \mathcal{T}_{{s_p}\left( {{l_p}} \right)}\right)\\
\label{32}
&{f_{{i{d_p}\left( {{l_p}} \right)}}^p}\left( l_p \right) \ge 0 \text{ }\text{ } \left(l_p \in \mathcal{L}_p, i \in \mathcal{R}_{{d_p}\left( {{l_p}} \right)}\right)\\
&{f_{{s_p}\left( {{l_p}} \right)j}}\left( l \right) = {f_{i{d_p}\left( {{l_p}} \right)}}\left( l \right) = 0  \nonumber \\  & \left(l \in \mathcal{L}, j \in \mathcal{T}_{{s_p}\left( {{l_p}} \right)}, i \in \mathcal{R}_{{d_p}\left( {{l_p}} \right)}, l_p \in \mathcal{L}_p\right) \\
&{f_{{s_p}\left( {{l_p}} \right)j}^p}\left( {{l_p}'} \right) = {f_{i{d_p}\left( {{l_p}} \right)}^p}\left( {{l_p}'} \right)= 0 \nonumber \\ &  \left({l_p} \ne {l_p}', {l_p}, {l_p}' \in \mathcal{L}_p \right) \\
\label{33}
&{\theta _{{l_p}}} \in \left\{ {0,1} \right\} \text{ }\text{ } \left(l_p \in \mathcal{L}_p\right) \text{ }\text{ } {\Upsilon _l} \ge 0 \text{ }\text{ }\text{ } \left(l \in \mathcal{L}\right)
\end{align}
where ${\theta _{{l_p}}}$, ${f_{ij}}\left( l \right)$, ${f_{ij}^p}\left( l_p \right)$, ${f_{{{s_p}\left( {{l_p}} \right)}j}^p}\left( l_p \right)$, ${f_{{i{d_p}\left( {{l_p}} \right)}}^p}\left( l_p \right)$, ${f_{{s_p}\left( {{l_p}} \right)j}}\left( l \right)$, ${f_{i{d_p}\left( {{l_p}} \right)}}\left( l \right)$, ${f_{{s_p}\left( {{l_p}} \right)j}^p}\left( {{l_p}'} \right)$, ${f_{i{d_p}\left( {{l_p}} \right)}^p}\left( {{l_p}'} \right)$,  $\lambda_{mq}$ and ${\Upsilon _l}$ are decision variables. Although indicator functions and set membership functions have been employed in this problem (e.g., constraints (13)), they become constants given the results of the MIS searching subproblem. Clearly, after reformulating this problem based on the results of the MIS searching subproblem, both the objective function and constraints of the reformulated optimization problem are linear. The only integer decision variables involved are those 0-1 variables $\theta_{l_p}$ which signifies the cooperating decision of the SSP on the $l_p$th primary session. Thus, given the results of the MIS searching subproblem, above optimization problem is a mixed integer linear programming (MILP) which is generally NP-hard. Fortunately, in the MILP part of our formulation, the integer variables are 0-1 variables resulted from the selection of primary sessions. Noticing that the number of primary sessions in the considered areas will be limited due to potentially mutual interference between them, the number of integer variables in the MILP part of our formulation is limited and thus the considered MILP can be solved by optimization softwares, such as CPLEX and lp\_solve, employing, for example, the classical branch-and-bound approach. Thus, the most difficult part of the optimization problem is to search for MISs in $G=(V,E)$, which will be introduced next.

\subsection{Augmented SIO-Based Algorithm for MIS Search}
Generally, finding all MISs of a conflict graph $G=(V,E)$ is NP-complete and is often encountered in multi-hop wireless networks \cite{Li2010,Pan2014}. When the size of $G=(V,E)$ is small, all MISs can be found via brute-force search. When the size of $G=(V,E)$ becomes large, the complexity of brute-force search is prohibitive so that it is impractical to find all MISs \cite{Li2010}. Recently, the computation of MISs in multi-hop wireless networks has been systematically studied in \cite{Li2010}. In this work, Li et al. point out that only a small set of MISs, i.e., critical MISs, are needed and scheduled by the optimal solution although $G=(V,E)$ has exponentially many MISs. In view of that, they developed an SIO-based method to intelligently compute a set of MISs such that critical MISs are covered as many as possible. As shown in \cite{Li2010}, the SIO-based method returns a set of MISs in polynomial time and outperforms the widely adopted random algorithms.

To find the critical MISs for the considered throughput maximization problem, the SSP needs to know the locations of the sources and destinations of primary sessions. We assume such information can be obtained from primary users (PUs) or their service providers. This assumption is made based on the following considerations. First, this work addresses problems in cooperative cognitive radio networks (CRNs) where certain level of cooperation and information exchange exist between primary networks and secondary networks. Second, as a service provider, the SSP will have more credibility than individual SUs, which will facilitate such information sharing with PUs. With such information, the SSP can construct the PU-related conflict graph based on which the critical MISs can be found. Once the SSP knows which primary sessions to cooperate with, it can employ the SIO-based method to identify a set of MISs where critical MIS are covered as many as possible. Unfortunately, the sources and destinations of the considered cross-layer optimization problem are not known in advance since the SSP needs to intelligently select primary sessions to cooperate with in order to maximize cooperation-incurred benefits. Clearly, different primary sessions will lead to different sources and destinations and thus different critical MISs. To address this challenge, we develop an heuristic algorithm, called the augmented scheduling index ordering based (SIO-based) algorithm, based on the SIO-based method so that critical MISs can be covered as many as possible. Once the SSP obtains the PU-related conflict graph based on the information shared by PUs, it can employ the augmented SIO-based algorithm as well as the information on the CCHN and primary sessions to find a set of MISs where critical MISs are covered as many as possible.

The augmented SIO-based algorithm, as shown in Algorithm 1, is developed based on the observation that the uncertainty of sources and destinations comes from the selection of primary sessions. The basic idea of the proposed algorithm is to compute a set of MISs for every possible combination of primary sessions and collect all these computed MISs to augment the set of MISs computed by the original SIO-based method. Specifically, for each choice of primary sessions, we will first eliminate the unselected primary sessions and the links which conflict with these primary sessions from the PU-related conflict graph and, then, run the SIO-based algorithm on the resulted graph to obtain a set of MISs of this graph. After that, the unselected primary sessions are added back to each of these MISs to obtain a set of MISs of the original PU-related conflict graph. Once such a set of MISs is obtained, it is combined with the previously computed sets of MISs. Following this procedure, we can obtain the augmented set of MISs after going through all possible choices of primary sessions. Finally, this augmented set of MISs is combined with that computed by the original SIO-based algorithm into a new set of MISs which will be used in the considered optimization problem for solution finding. Given $L_p$ primary sessions, line $5-8$ will be iterated for $2^{L_p}-1$ times. As proved in \cite{Li2010}, the running time of line 6 is ${\rm O}\left( V^4 \right)$ and dominates the running time of each iteration. Noticing line $1$ takes ${\rm O}\left( V^4 \right)$ time, the complexity of the proposed algorithm is ${\rm O}\left( {{2^{{L_p}}}{V^4}} \right)$. Due to mutual conflict/interference, the number of primary sessions $L_p$ in a certain area is limited and is bounded by a constant. Then, the complexity of the proposed algorithm becomes ${\rm O}\left( {{V^4}} \right)$, which implies the proposed algorithm will terminate in polynomial time.

\begin{center}
\begin{algorithm}[!t]
\caption{: Augmented SIO-Based Algorithm}
\begin{algorithmic}[1]
\REQUIRE The topology of the CCHN, sources and destinations of primary sessions, and the PU-related conflict graph $G=(V,E)$,
\ENSURE A set of MISs $\mathcal{I}_a$
\STATE Compute a set of MISs $\mathcal{I}_a$ of $G$ based on the SIO-based method
\FOR{$j$=0 to $L_p-1$}
  \STATE Compute all subsets of $\mathcal{L}_p$ with cardinality $j$ and collect these subsets in a set $\mathcal{P}_j$;
  \FOR{all $p \in \mathcal{P}_j$}
  \STATE Construct another graph $G_p$ from the PU-related conflict graph $G$ by removing the primary sessions in $\mathcal{L}_p - p$ as well as the vertices/links conflicting with these primary sessions
  \STATE Compute a set of MISs of $G_p$, $\mathcal{M}_p$, with $\left\{{s(l), l \in \mathcal{L}}\right\} \cup \left\{{s(l_p), l_p \in p}\right\}$ as the source and $\{b\} \cup \left\{d_p(l_p), l_p \in p\right\}$ as the destinations based on the SIO-based method
  \STATE add the primary session in $\mathcal{L}_p - p$ to each set in $\mathcal{M}_p$ to obtain a set, $\mathcal{I}_p$, of MISs in $G$
  \STATE $\mathcal{I}_a$=$\mathcal{I}_a \cup \mathcal{I}_p$
  \ENDFOR
\ENDFOR
\end{algorithmic}
\end{algorithm}
\end{center}

\section{Performance Evaluation}
In this section, the feasibility and effectiveness of the proposed NLC scheme is examined via extensive simulations.

\subsection{Simulation Setup}
\begin{figure}[!t]
 \begin{center}
  \includegraphics[width=3in]{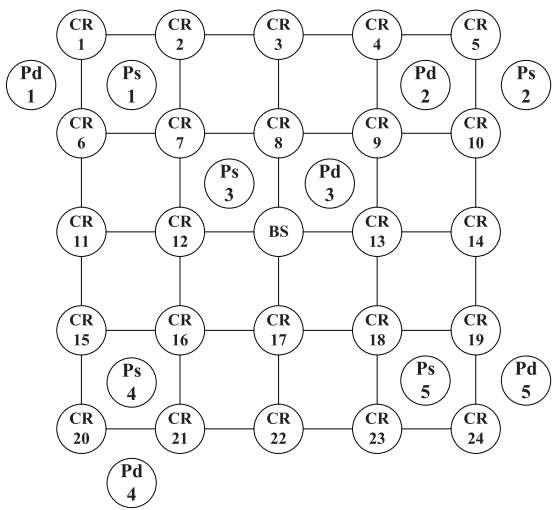}
  \end{center}
  \begin{center}
   \parbox{6cm}{\caption{Topology of the simulated CCHN.}  \label{fig5}}
  \end{center}
  \end{figure}

We consider a CCHN with a BS and $\mathcal{N}=24$ CR routers. According to \cite{Yue2013}, the placement of CR routers should be carefully planned to improve the spectrum efficiency and system capacity. As a result, we assume the BS and CR routers are regularly deployed based on a grid topology as shown in Fig. \ref{fig5} where the BS is located at the center and each pair of secondary network facilities is $200m$ away. Among those CR router, $\text{CR}1$ and $\text{CR}24$ are edge CR routers. There are $5$ primary sessions collocated with the CCHN and the source of each session is $200m$ far away from its destination. The sources and destinations of primary sessions are $100\sqrt{2}m$ away from the nearest secondary network facilities. The BS, CR routers and the source of each primary session all employ $2 W$ for transmission, i.e., $P_t^{\mu}=2W$, $\forall \mu \in \left\{ {C,b,P} \right\}$. The thresholds for successful reception and the interference thresholds are set as $P_R^{\nu}=10^{-6} W$ and $P_I^{\nu}=1.34 \times 10^{-7} W$, $\forall \nu \in \left\{ {C,b,P} \right\}$, respectively. The path loss exponent is $n=3$ and the antenna related constant $\gamma=4.63$. Based on Section IV.B, $R_T^{\mu \nu}=210 m$ and $R_I^{\mu \nu}=410 m$, $\forall \mu, \nu \in \left\{ {C,b,P} \right\}$, i.e., CR routers, the BS, and PUs share the same transmission range and interference range.

\subsection{Results and Analysis}
\begin{figure}[!t]
 \begin{center}
  \includegraphics[width=3in]{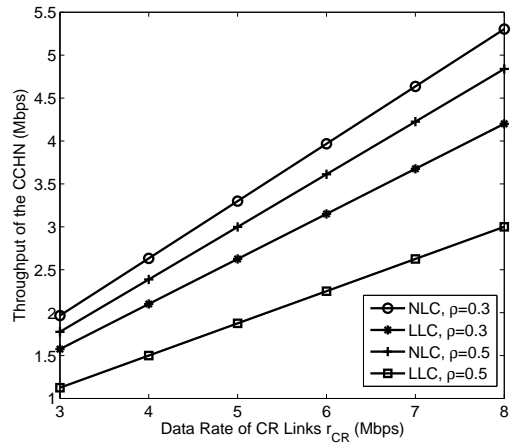}
  \end{center}
  \begin{center}
   \parbox{8cm}{\caption{The throughput performance of the NLC approach and the LLC approach.}  \label{fig10}}
  \end{center}
  \end{figure}

The performance of our NLC approach is first compared with that of the LLC approach in Fig. \ref{fig10} where the LLC approach is implemented based on decode-and-forward relaying with a frame length of $10ms$ \cite{Cao2013}. The PU-related links have the same data rate $3Mbps$, the CR links have the same data rate $r_{CR}$. To make our comparison more comprehensive, we introduce $\rho$, the probability that PUs are active, to signify PUs' activity and obtain final results by averaging the corresponding throughput of the CCHN when PU are active and inactive. In the case where PUs are active, the data volume of primary sessions is set as $D_1=\cdots=D_{L_p}=20Mbits$, and the lengths of the $5$ primary sessions are $30s, 30s, 30s, 60s, 60s$, which implies that the length of the control interval is $T=30s$. For the LLC approach, we assume PUs equally allocate their scheduled data into different frames. For our NLC approach, we set the incentive parameter $\alpha$ as $1$. As shown in Fig. \ref{fig10}, our NLC approach can achieve much higher throughput than that of the LLC approach, which demonstrates the effectiveness of our approach in network throughput enhancement. With a higher $r_{CR}$, the CCHN is able to deliver more data during a fixed time period, and thus it is not surprising that the throughput of the CCHN grows with $r_{CR}$ increasing. Another observation from Fig. \ref{fig10} is that the throughput of the CCHN decreases when $\rho$ increases from $0.3$ to $0.5$. Intuitively, the increase in PUs' activities will limit the number of network resources available to the CCHN, which will in turn lead to the reduction in the throughput of the CCHN.

\begin{figure}[!t]
 \begin{center}
  \includegraphics[width=3in]{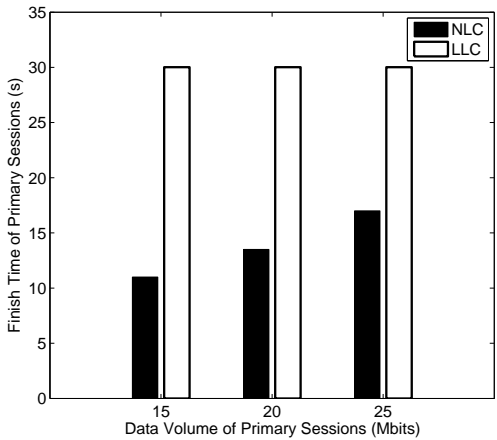}
  \end{center}
  \begin{center}
   \parbox{8cm}{\caption{The completion time of primary transmissions under the NLC approach and the LLC approach.}  \label{fig11}}
  \end{center}
  \end{figure}

In Fig. \ref{fig11}, we further compare the completion time of primary transmissions under our NLC approach with that under the LLC approach. To facilitate the comparison, we focus on the data transmission of the first primary session, i.e., the data transmission from {\rm Ps}1 to {\rm Pd}1, and assume it is active with probability 1. The LLC approach is implemented based on decode-and-forward relaying \cite{Cao2013}. Other parameters are the same as those in Fig. \ref{fig10}. As shown in Fig. \ref{fig11}, when compared with the LLC approach, our NLC approach can greatly shorten the completion time of primary session and thus are more likely to motivate PUs to join the cooperation-based spectrum access processes. From Fig. \ref{fig11}, with the data volume of the primary session growing, the completion time of the primary transmission increases when our NLC approach is adopted and almost remains the same when the LLC approach is employed. Clearly, given the network topology of the CCHN and the amount of available spectrum, the completion time of the primary transmission will increase when the data volume of the primary session increases, which explains the results under our NLC approach. As aforementioned, under the LLC approach, no matter how fast PUs' data could be delivered in each frame, PUs still need to wait until the last frame for their data transmissions to be finished. Due to small frame length, under the LLC scheme, the completion time of the primary transmission is almost the same when the volume of primary session varies.

\begin{figure}[!t]
 \begin{center}
  \includegraphics[width=3in]{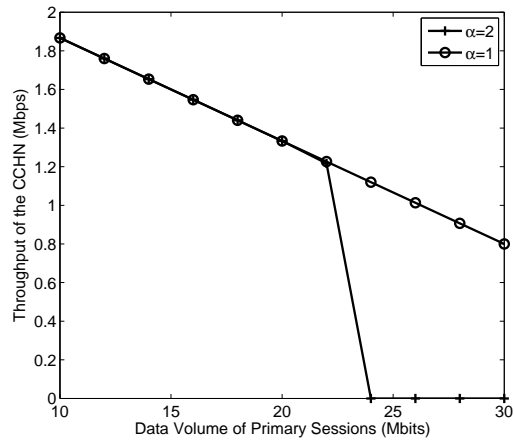}
  \end{center}
  \begin{center}
   \parbox{8cm}{\caption{The throughput of the CCHN v.s. the data volume of primary sessions.}  \label{fig6}}
  \end{center}
  \end{figure}

In Fig. \ref{fig6}, we study the relation between the throughput of the CCHN and the data volumes of primary sessions $D_{l_p}$'s. To clearly reflect the impact of $D_{l_p}$'s, we set $D_1=\cdots=D_{L_p}=D$, $\rho=1$, and assume both CR links and PU-related links have the same data rate $3Mbps$. The lengths of primary sessions are the same as those in Fig. \ref{fig10}. Fig. \ref{fig6} shows the throughput of the CCHN decreases as $D$ increases. During a control interval, the number of available network resources in the CCHN is fixed once the SSP determines which primary sessions to cooperate with. When $D$ gets larger, the SSP will allocate more resources to relay PUs' traffic and less resources will be used to deliver secondary data, which leads to a reduction in the amount of delivered secondary data during the control interval. Consequently, the growth of $D$ results in a decrease of the throughput of the CCHN. Additionally, the impact of the incentive parameter $\alpha$ shown in Fig. \ref{fig6} is very interesting. When the data volume $D$ of the primary sessions is small, the CCHN can obtain the same throughput under $\alpha=1$ and $\alpha=2$. However, after $D$ reaches a certain value, the throughput of the CCHN under $\alpha=2$ becomes $0$. In the considered network, when $D$ is small, delivering PUs' data will not cost too much and the SSP will choose to cooperate with those primary sessions no matter $\alpha=1$ or $\alpha=2$. Generally, the amount of available network resources is fixed once the SSP decides which primary sessions to cooperate with. Given the same amount of primary data traffic, the amount of network resources left for secondary flows is the same for $\alpha=1$ and $\alpha=2$ cases, which results in the same throughput of the CCHN under $\alpha=1$ and $\alpha=2$. When $D$ is large enough, things become different since PUs' data must be delivered in a shorter period of time when $\alpha=2$. In this case, the requirements of primary sessions are too high to be satisfied and thus the SSP chooses not to cooperate, which leads to a $0$ throughput.

\begin{figure}[!t]
 \begin{center}
  \includegraphics[width=3in]{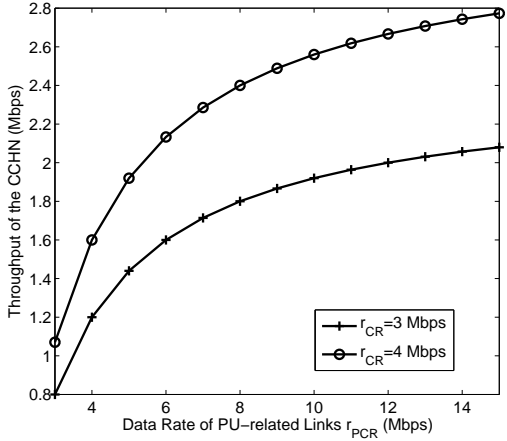}
  \end{center}
  \begin{center}
   \parbox{8cm}{\caption{The throughput of the CCHN v.s. the data rate of PU-related links.}  \label{fig7}}
  \end{center}
  \end{figure}

How the data rates of different links affect the throughput of the CCHN is shown in Fig. \ref{fig7}. In general, the SSP is able to schedule two kinds of links, PU-related links and CR links. To study the impacts of these links, we assume all PU-related links have the same data rate $r_{PCR}$ (i.e., $r_{s_p\left(l_p\right)j}=r_{id_p\left(l_p\right)}=r_{PCR}$, $i,j \in \mathcal{N}_s$, $l_p \in \mathcal{L}_p$) and all CR links have the same data rate $r_{CR}$ (i.e., $r_{ij}=r_{CR}$, $i,j \in \mathcal{N}_s$). The other parameters are the same as in Fig. \ref{fig6}, except $\alpha=1$ and $D=30Mbits$. It is observed that the throughput of the CCHN grows at a decreasing growth rate with $r_{PCR}$ increasing. When $r_{PCR}$ is higher, the CCHN can help PUs finish their transmission more quickly and obtain longer cooperation-incurred periods to deliver more secondary data. As a result, the CCHN obtains higher throughput with $r_{PCR}$ increasing. When $r_{PCR}$ is high enough, further increases in $r_{PCR}$ will not extend cooperation-incurred periods too much and thus the growth rate decreases. Since the secondary flows are carried by CR links, high-speed CR links will result in improvement in the throughput of the CCHN as shown in Fig. \ref{fig7}. Particularly, when cooperation-incurred periods are extended due to high $r_{PCR}$, much more secondary data can be delivered with higher $r_{CR}$, which explains the gap between the two curves in Fig. \ref{fig7}.

\begin{figure}[!t]
 \begin{center}
  \includegraphics[width=3in]{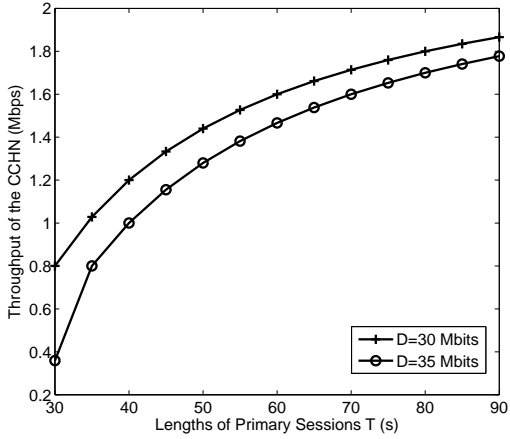}
  \end{center}
  \begin{center}
   \parbox{8cm}{\caption{The throughput of the CCHN v.s. the lengths of primary sessions.}  \label{fig8}}
  \end{center}
  \end{figure}

In Fig. \ref{fig8}, we study the relationship between the throughput of the CCHN and the lengths of primary sessions. To make it more clear, we assume all primary sessions have the same length, i.e., $T_1=\cdots=T_{L_p}$, and thus the length of control interval is $T=T_1=\cdots=T_{L_p}$. The values of other parameters are the same as in Fig. \ref{fig6}. The results show the throughput of the CCHN is an increasing function of $T$ with decreasing growth rate. Let $\tau_m$ be the maximum achievable throughput of the CCHN when all network resources are dedicated to secondary data transmissions, $\tau_T$ and $\tau_{T+\Delta T}$ be the achievable throughput of the CCHN via cooperating with primary sessions over control intervals with lengths of $T$ and $T+\Delta T$, respectively. Intuitively, $\tau_m \ge \tau_T$. For illustration, we assume the CCHN can deliver PUs' data in $T$ and the SSP's cooperating decision on each primary session remains unchanged when $T$ is extended to $T+\Delta T$. As a result, during the period $\Delta T$, all network resources will be used for secondary transmissions. Then, we have ${\tau _{T{\rm{ + }}\Delta T}}{\rm{ = }}\frac{{{\tau _T}T + {\tau _m}\Delta T}}{{T{\rm{ + }}\Delta T}} \ge \frac{{{\tau _T}T + {\tau _T}\Delta T}}{{T{\rm{ + }}\Delta T}} = {\tau _T}$, which explains why the throughput of the CCHN increases when $T$ is extended to $T+\Delta T$. Additionally, the growth rate of the throughput can be derived as $\frac{{{\tau _{T{\rm{ + }}\Delta T}} - {\tau _T}}}{{\Delta T}} = \frac{{\left( {{\tau _m} - {\tau _T}} \right)}}{{T{\rm{ + }}\Delta T}}$. Since $\tau_T$ increases with respect to $T$, the growth rate of $\tau_T$, i.e., $\frac{{{\tau _{T{\rm{ + }}\Delta T}} - {\tau _T}}}{{\Delta T}}$, decreases when $T$ gets larger.

\begin{figure}[!t]
 \begin{center}
  \includegraphics[width=3in]{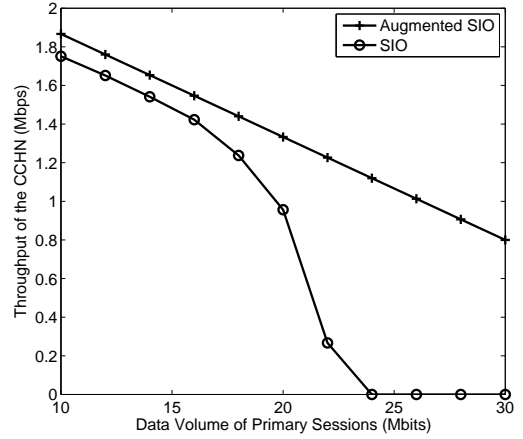}
  \end{center}
  \begin{center}
   \parbox{8cm}{\caption{Comparison between the augmented SIO-based algorithm and the original SIO-based method.}  \label{fig9}}
  \end{center}
  \end{figure}

To examine the performance of the augmented SIO-based algorithm, we compare the maximum throughput of the CCHN based on the augmented SIO-based algorithm with that based on the original SIO-based method in Fig. \ref{fig9}. The parameter settings are the same as Fig. \ref{fig6} except the incentive parameter $\alpha=1$. The results demonstrate the superiority of the augmented SIO-based algorithm, particularly when primary sessions have a large amount of data to transmit. According to Fig. \ref{fig9}, the original SIO-based method can achieve comparable performance with that of the augmented SIO-based method when $D$ is small. Intuitively, a small $D$ means PUs' data can be easily delivered and the maximum throughput of the CCHN mainly depends on the scheduling of secondary flows. In the CCHN, the sources and the destinations of secondary flows are known to be the edge CR-routers and the BS, which implies the SIO-based method can cover most of the critical MISs for secondary flows. Consequently, when $D$ is small, based on the SIO-based method, the achievable throughput is close to that based on the augmented SIO-based method. When $D$ gets larger, the SSP will allocate more resources to relay PUs' traffic to acquire the spectrum access opportunities. In this case, the achievable throughput of the CCHN will not only be determined by the scheduling of secondary flows but also be affected by how the primary traffic is delivered. Since the SIO-based method is not efficient in computing the critical MISs for PUs' data delivery, the CCHN will get lower throughput by scheduling the set of MISs computed from the original SIO-based method.

 \begin{figure}[!t]
 \begin{center}
  \includegraphics[width=3in]{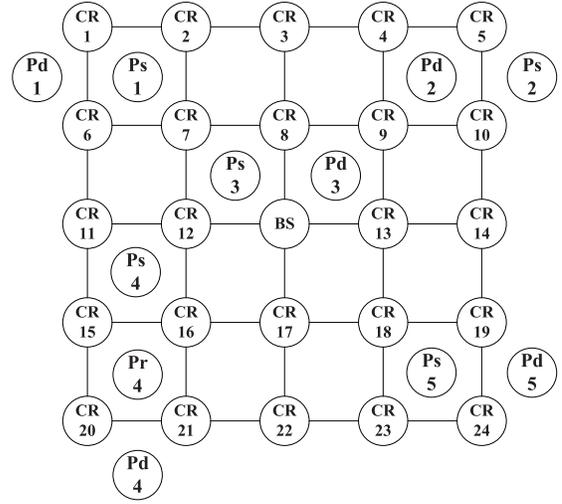}
  \end{center}
  \begin{center}
   \parbox{8cm}{\caption{Topology of the simulated CCHN with some primary sessions implemented via multi-hop transmissions.}  \label{figr1}}
  \end{center}
  \end{figure}

\begin{figure}[!t]
 \begin{center}
  \includegraphics[width=3in]{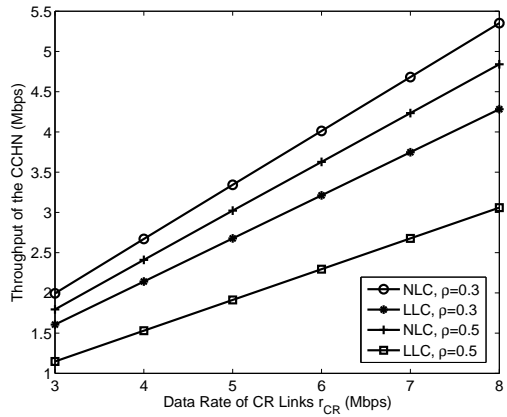}
  \end{center}
  \begin{center}
   \parbox{8cm}{\caption{The throughput performance of the NLC approach and the LLC approach, when some primary sessions are implemented via multi-hop transmissions.}  \label{figr2}}
  \end{center}
  \end{figure}

\begin{figure}[!t]
 \begin{center}
  \includegraphics[width=3in]{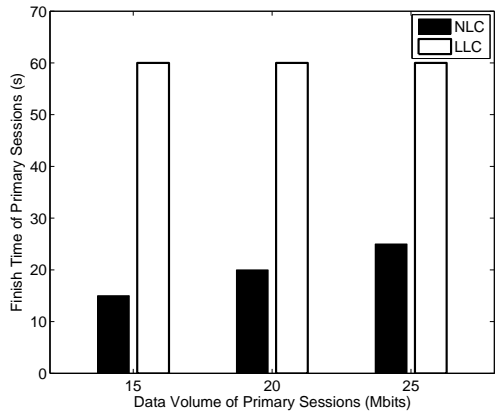}
  \end{center}
  \begin{center}
   \parbox{8cm}{\caption{The completion time of primary transmissions, under the NLC scheme and the LLC scheme, when some primary sessions are implemented via multi-hop transmissions.}  \label{figr3}}
  \end{center}
  \end{figure}

Finally, we evaluate the effectiveness of our NLC scheme when some primary sessions are implemented via multi-hop transmissions. Specifically, we consider a CCHN where the fourth primary session, i.e., the data transmission from {\rm Ps}4 to {\rm Pd}4, is implemented via multi-hop transmissions and {\rm Pr}4 is the intermediate relay for the fourth primary session as shown in Fig. \ref{figr1}. The parameters and network settings are the same as those in Fig. \ref{fig10}. The only difference is that, in each frame, the transmission of the fourth primary session is completed over two subframes of length $5ms$ and the LLC scheme is implemented in each subframe through decode-and-forward relaying since there are two hops involved in the fourth primary session. The results are shown in Fig. \ref{figr2}. It is obvious that our NLC scheme can achieve a much higher throughput than that of the LLC approach, which implies that our NLC approach is effective in network throughput enhancement even if multi-hop flows are involved in the primary networks. Based on the same network topology, we further compare the completion time of primary sessions under our NLC approach and that under the LLC approach in Fig. \ref{figr3}. To examine the effectiveness of our NLC approach in dealing with multi-hop primary flows, we focus on the data transmission of the fourth primary session and assume it is active with probability $1$. Other parameters are the same as those in Fig. \ref{figr2}. Clearly from Fig. \ref{figr3}, even if primary sessions are implemented via multi-hop transmissions, our NLC approach can still greatly shorten the completion time of primary sessions when compared with the LLC approach and thus is more likely to motivate PUs to join the cooperation-based spectrum access processes.

\section{Throughput Analysis}
Previous sections are dedicated to the NLC mechanism design under the CCHN architecture, which focuses on the achievable throughput of the CCHN. In this section, we will analyze what our CCHN based NLC scheme can offer to individual SUs.

Specifically, we consider a network where $n$ SUs distribute in a square with side length $n^{\frac{1}{2}}$. There are totally $n^b$ BSs and CR routers regularly placed in the considered square among which there are $n^d$ BSs, where $0 < d \le b < 1$. As proved in Appendix B, the achievable throughput of individual SUs can be derived as
\begin{align}
\label{r26}
\xi  = \left\{ {\begin{array}{*{20}{c}}
{\begin{array}{*{20}{c}}
{\Omega \left( {{n^{b - 1}}} \right)}&{0 < d = b < 1}
\end{array}}\\
{\begin{array}{*{20}{c}}
{\Omega \left( {\min \left\{ {{n^{b - 1}},W{n^{\frac{b}{2} - 1}}} \right\}} \right)}&{0 < \frac{b}{2} < d < b < 1}
\end{array}}\\
{\begin{array}{*{20}{c}}
{\Omega \left( {\min \left\{ {{n^{b - 1}},W{n^{d - 1}}} \right\}} \right)}&{0 < d \le \frac{b}{2} < 1}
\end{array}}
\end{array}} \right.,
\end{align}
where $W$ is the bandwidth which can be exploited by BSs and CR routers for data transmissions and is obtained by, for example, helping with PUs¡¯ transmissions. $0<b=d<1$ corresponds to the case where all nodes deployed by the SSP are BSs.

From (\ref{r26}), we can gain a couple of insights on the behaviour of the CCHN when the number of SUs is high. Clearly from (\ref{r26}), the achievable throughput of each SU decreases when more SUs are served by the CCHN. This is not surprising since there will be more SUs contending for resources. Fortunately, as shown in (\ref{r26}), depending on specific situations, the SSP can improve its service provisioning via either identifying more spectrum resources or deploying more BSs/CR routers. For example, given ${0 < \frac{b}{2} < d < 1}$ and the number of BSs (i.e., $d$), the achievable throughput of individual SUs is determined by $b$ and $W$ which represents the number of CR routers and the number of harvested spectrum resources available to BSs/CR routers. In this case, the SSP can offer higher achievable throughput by deploying more CR routers and acquiring more harvested bands. Once enough CR routers are deployed, i.e., ${0 < d \le \frac{b}{2} < 1}$, the achievable throughput of SUs is limited by the number of BSs due to contention at BSs. In this case, the SSP should deploy more BSs in order to resolve contention. Deploying extra BSs and CR routers will not only result in an increase in both the OPEX and CAPEX of the SSP but also increase the complexity of network management. Thus, the deployment of BSs and CR routers in the CCHN should be carefully studied. That is the reason why we study the placement of BSs and CR routers in \cite{Yue2013}.

\section{Conclusion}
In this paper, we propose a network-level session-based cooperative (NLC) approach to enable the cooperation-based spectrum access in CRNs. Unlike the traditional link-level frame-based cooperative (LLC) approach, in the NLC approach, a group of SUs instead of individual SUs cooperate with primary sessions for spectrum access opportunities and the obtained spectrum access opportunities are sharing within the corresponding group. Thanks to the group-based cooperation approach and session-by-session cooperative strategy, the NLC approach is able to achieve efficient spectrum resource utilization for network capacity enhancement and motivate PUs to join the cooperation-based spectrum access processes. To elaborate our NLC approach, we further develop an NLC scheme based on our CCHN architecture. By leveraging the PU-related conflict graph, we mathematically formulate the cooperative mechanism design as a throughput maximization problem with constraints on primary session selection, flow routing, and link scheduling. To facilitate the cross-layer optimization, we develop an augmented algorithm based on scheduling index ordering (SIO) to search for MISs. The impacts of various network parameters on the throughput of the CCHN are carefully studied via extensive simulations. Our extensive simulation studies demonstrate the superiority of the proposed CCHN based NLC scheme which provides a promising solution for future cooperative CRNs.

\appendices
\section{The Decomposable Structure of the Considered Problem}
The optimization problem studied in this paper aims to maximize the throughput of the CCHN under interference constraints. For efficient resource utilization, we need to jointly consider the flow routing in the network layer and the link scheduling in the data link layer. In essence, the optimization problem considered in this paper is a throughput maximization problem for multihop wireless networks with interference constraints. Generally, this kind of problems can be formulated as
\begin{align}
\label{r1}
 &{\rm{maximize    }}\sum\limits_{l \in \mathcal{L}} {{\Upsilon _l}}  \nonumber \\
 s.t.: \text{ }\text{ }\text{ }&\text{flow} \text{ } \text{routing} \text{ } \text{constraints},  \\
 &{\text{ }\text{ }}\overline {{f_{ij}}}  \le {v_{ij}}{c_{ij}}, 0 \le v_{ij} \le 1,\nonumber\\
 &\text{Interference}\text{ }\text{constraints}\text{ }\text{for}\text{ } v_{ij} \nonumber,
\end{align}
where the objective function is the maximum throughput supported over the CCHN, $\overline {{f_{ij}}}$ is the total amount of flow allocated to link $(i,j)$, $c_{ij}$ is the throughput of link $(i,j)$, $v_{ij}$ is the time share where link $(i,j)$ is active. The ``Interference constraints for $v_{ij}$'' in (C1) signifies the interference relationships between different links, such as two links cannot be active simultaneously, which limit the value of $v_{ij}$. In this paper, such interference relationships are characterized by the conflict graph $G$. The decision variables for problem (C1) is $\overline{f_{ij}}$ and $v_{ij}$ which are collected in vectors $\textbf{f}$ and $\textbf{v}$, respectively. As proved in \cite{Jain2003}, it is not only NP-hard to find the optimal solution to problem (C1) but also NP-hard to solve it approximately, which implies that we should look for heuristic algorithms.

As shown in \cite{Jain2003}, $\bf{v}$ satisfies the ``Interference constraints for $v_{ij}$'' if and only if it is within the independent set polytope of the conflict graph $G$. Namely, $\bf{v}$ is feasible under the ``Interference constraints for $v_{ij}$'' if and only if
\begin{align}
\label{r2}
\textbf{v}=\sum\limits_{\tau = 1}^{\mathcal Q} {{\lambda _\tau}{\textbf{v}_\tau}}, \sum\limits_{\tau = 1}^{\mathcal Q} {{\lambda _\tau}}  = 1, 0\le \lambda_\tau \le 1,
\end{align}
where ${\textbf{v}_\tau} \in {\left\{ {0,1} \right\}^{\left| V \right| \times 1}}$ is a vector used to characterize the $\tau$th independent set of the conflict graph $G$, $\left| V \right|$ is the number of vertices in conflict graph $G$. The $j$th element of ${\textbf{v}_\tau}$ is $1$ if and only if the corresponding vertex is in the $\tau$th independent set. $\mathcal Q$ is the total number of independent sets of $G$. By replacing the ``Interference constraints for $v_{ij}$'' in (\ref{r1}) with (\ref{r2}), we have an equivalent optimization problem as
\begin{align}
\label{r3}
 &{\rm{maximize    }}\sum\limits_{l \in \mathcal{L}} {{\Upsilon _l}}  \nonumber \\
 s.t.: \text{ }\text{ }\text{ }&\text{flow} \text{ } \text{routing} \text{ } \text{constraints},  \\
 &\overline {{f_{ij}}}  \le {c_{ij}}\sum\limits_{\tau = 1}^{\mathcal Q} {{\lambda _\tau}1\left( {\left( {i,j} \right) \in {{\rm I}_\tau}} \right)},\nonumber\\
 &\sum\limits_{\tau = 1}^{\mathcal Q} {{\lambda _\tau}}  = 1, 0\le \lambda_\tau \le 1 \nonumber,
\end{align}
where ${\rm I}_\tau$ represents the $\tau$th independent set of $G$ and ${1\left( {\left( {i,j} \right) \in {{\rm I}_\tau}} \right)}$ is an indicator function which equals $1$ if and only if the vertex associated with link $(i,j)$ is within ${\rm I}_\tau$. $\lambda_\tau$ can be viewed as the time share allocated to ${\rm I}_\tau$. For ease of presentation, we collect $\lambda_\tau$'s in a vector $\Lambda$.

According to the definition of the maximal independent set (MIS), each independent set must be a subset of an MIS. By offering the time share allocated to ${\rm I}_\tau$ to the MIS $I_q$, where ${\rm I}_\tau \subseteq I_q$ and $I_q$ is the $q$th MIS, we can always obtain a feasible solution $\Lambda'$ to problem (\ref{r3}) from each feasible solution $\Lambda$ to problem (\ref{r3}) and achieve at least the same maximum achievable throughput. Thus, to find the optimal solution to problem (\ref{r3}), we only need to consider $\Lambda$ with $\lambda_\tau=0$ if the $\tau$th independent set, ${\rm I}_\tau$, is not an MIS of $G$. Namely, we can find an optimal solution to problem (\ref{r3}) by solving the following optimization problem
\begin{align}
\label{r4}
&{\rm{maximize    }}\sum\limits_{l \in \mathcal{L}} {{\Upsilon _l}}  \nonumber \\
 s.t.: \text{ }\text{ }\text{ }&\text{flow} \text{ } \text{routing} \text{ } \text{constraints},  \\
 &\overline {{f_{ij}}}  \le {c_{ij}}\sum\limits_{q = 1}^Q {{\lambda _q}1\left( {\left( {i,j} \right) \in {{I}_q}} \right)},\nonumber\\
 &\sum\limits_{q = 1}^{Q} {{\lambda _q}}  \le 1, 0\le \lambda_q \le 1 \nonumber,
\end{align}
where $Q$ is the total number of MISs of $G$ and $I_q$ is the $q$th MIS of $G$. $\sum\limits_{q = 1}^{Q} {{\lambda _q}}  \le 1$ is due to the fact that empty set is an independent set but not an MIS. Then, we can achieve the optimal throughput and obtain the optimal solution to problem (\ref{r3}) by solving problem (\ref{r4}). Obviously, given all MISs of $G$, problem (\ref{r4}) is a linear programming if no integer variables are involved in the flow routing constraints, which implies that problem (\ref{r4}) and thus the original problem (\ref{r1}) can be solved by solving two subproblems, i.e., finding all MISs of $G$ and the corresponding linear programming to find the optimal scheduling of these MISs as well as flow allocation. The only difference between our problem and problem (\ref{r1}) is that our problem involves the selection of primary sessions which not only introduces 0-1 variables to flow routing constraints but also affect the set of MISs to be scheduled. Due to the involvement of these 0-1 variables, our problem can be decomposed into an MIS searching subproblem and an MILP, instead of a linear programming.

\section{Achievable Throughput of Individual SUs}
In this paper, our proposed NLC scheme is studied based on the CCHN where an SSP deploys BSs and CR routers to serve SUs by harvesting spectrum resources. In the CCHN, SUs' data is either aggregated at neighboring edge CR routers or directly delivered to BSs, if possible. Under the supervision of the SSP, CR routers collectively exploit harvested spectrum resources, such as those obtained via helping with PUs' transmissions, and deliver SUs' data to BSs for data networks access. Obviously, data transmissions in the CCHN are carried out in at most two steps: from SUs to neighboring BSs/CR routers and from CR routers to targeted BSs. Thus, we will analyze SUs' achievable throughput in the two steps, respectively, and the minimum of these two is SUs' achievable throughput in the CCHN. To facilitate analysis, we make the following assumptions
\begin{itemize}
  \item All SUs have a total bandwidth of 1 which can be utilized to access neighboring BSs/CR routers. CR routers and BSs can exploit a bandwidth of $W$ which is obtained by, for example, helping with PUs' transmissions.
  \item There are $n$ SUs distributed in a square with side length $n^{1/2}$. There are totally $m=n^b$ BSs and CR routers in the considered square among which there are $n^d$ BSs, where $0<d \le b<1$. These BSs and CR routers are regularly places so that the consider square can be divided into squares of area $\frac{1}{m}$, which we call subsquares. Clearly, there is only one BS or CR router in each subsquare.
  \item BSs are evenly distributed along both the $y$-coordinate and the $x$-coordinate. As a result, for the subsquares sharing the same $y$-coordinates, when $d > \frac{b}{2}$, there are $n^{d-\frac{b}{2}}$ of them containing a BS, when $d \le \frac{b}{2}$, there are at most $c_1$ of them containing a BS, where $c_1$ is a constant. Meanwhile, for the subsquares sharing the same $x$-coordinates, when $d > \frac{b}{2}$, there are $n^{d-\frac{b}{2}}$ of them containing a BS, when $d \le \frac{b}{2}$, there are at most $c_2$ of them containing a BS, where $c_2$ is a constant.
  \item Each SU will generate a flow which will be first aggregated at the BS or CR router within the same subsquare. If a BS resides in a subsquare, it is considered as the destination of all flows generated from this subsquare. While, for CR routers, flows generated from the corresponding subsquares are assumed to be targeted at randomly selected BSs, for easy of analysis.
  \item SUs can reach the BSs/CR routers in the same subsquare via either direct transmissions or multi-hop transmissions depending on their distances.
  \item The inter-subsquare transmissions are carried out between corresponding BSs/CR routers.
\end{itemize}

Before we proceed, we will prove a Lemma on the probability distribution of a random variable $X=\sum\limits_{i = 1}^\eta  {{\varepsilon _i}}$, where $\varepsilon_i$'s are mutually independent Bernoulli random variables, and $\varepsilon_i$ equals $1$ with probability $p_i$. Noticing that this Lemma is provided in \cite{Panli2012} without proof, we offer the proof here to facilitate easy reading.

\begin{lemma}
For any $\delta>0$, we have
\begin{align}
\label{r8}
{\rm P}\left( {X \ge \left( {1 + \delta } \right){\mu _X}} \right) < {e^{ - {\mu _X}f\left( \delta  \right)}},
\end{align}
where $\mu_X$ is the mean of $X$ and $f\left( \delta  \right) = \left( {1 + \delta } \right)\ln \left( {1 + \delta } \right) - \delta$.
\end{lemma}
\begin{IEEEproof}
For $r>0$ and $\sigma>0$, we have
\begin{align}
\label{r9}
{\rm P}\left( {X - {\mu _X} \ge r} \right) = {\rm P}\left( {{e^{\sigma \left( {X - {\mu _X}} \right)}} \ge {e^{\sigma r}}} \right).
\end{align}
Applying Markov inequality, the probability in (\ref{r9}) can be bounded as
\begin{align}
\label{r10}
{\rm P}\left( {X - {\mu _X} \ge r} \right) \le {\rm E}\left[ {{e^{\sigma \left( {X - {\mu _X}} \right)}}} \right]{e^{ - \sigma r}}.
\end{align}
Noticing that $X=\sum\limits_{i = 1}^\eta  {{\varepsilon _i}}$ and ${\mu _X} = \sum\limits_{i = 1}^\eta  {{\mu _{{\varepsilon _i}}}}$, where ${{\mu _{{\varepsilon _i}}}}$ is the mean of ${{\varepsilon _i}}$ and equals $p_i$, it follows
\begin{align}
\label{r11}
{\rm E}\left[ {{e^{\sigma \left( {X - {\mu _X}} \right)}}} \right] =& {\rm E}\left[ {{e^{\sigma \sum\limits_{i = 1}^\eta  {{\varepsilon _i} - {\mu _{{\varepsilon _i}}}} }}} \right] = {e^{ - \sigma \sum\limits_{i = 1}^\eta  {{\mu _{{\varepsilon _i}}}} }}{\rm E}\left[ {{e^{\sigma \sum\limits_{i = 1}^\eta  {{\varepsilon _i}} }}} \right] \nonumber \\=& {e^{ - \sigma {\mu _X}}}{\rm E}\left[ {{e^{\sigma \sum\limits_{i = 1}^\eta  {{\varepsilon _i}} }}} \right]= {e^{ - \sigma {\mu _X}}}\prod\limits_{i = 1}^\eta  {{\rm E}\left[ {{e^{\sigma {\varepsilon _i}}}} \right]} \nonumber \\=& {e^{ - \sigma {\mu _X}}}\prod\limits_{i = 1}^\eta  {\left( {{p_i}{e^{\sigma}} + 1 - {p_i}} \right)}.
\end{align}

Since ${p_i}{e^{\sigma}} - {p_i}>0$ and ${e^x} > 1 + x$, $\forall x>0$, we have
\begin{align}
\label{r12}
{p_i}{e^\sigma } + 1 - {p_i} < {e^{{p_i}{e^\sigma } - {p_i}}}.
\end{align}

Plugging (\ref{r12}) into (\ref{r11}), it follows
\begin{align}
\label{r13}
{\rm E}\left[ {{e^{\sigma \left( {X - {\mu _X}} \right)}}} \right] <& {e^{ - \sigma {\mu _X}}}\prod\limits_{i = 1}^\eta  {{e^{{p_i}{e^\sigma } - {p_i}}}}  = {e^{ - \sigma {\mu _X}}}{e^{\sum\limits_{i = 1}^\eta  {{p_i}{e^\sigma } - {p_i}} }} \nonumber \\=& {e^{ - \sigma {\mu _X} - {\mu _X} + {\mu _X}{e^\sigma }}}.
\end{align}
With (\ref{r10}) and (\ref{r13}), ${\rm P}\left( {X - {\mu _X} \ge r} \right)$ can be further bounded as
\begin{align}
\label{r14}
{\rm P}\left( {X - {\mu _X} \ge r} \right) < {e^{ - \sigma r - \sigma {\mu _X} - {\mu _X} + {\mu _X}{e^\sigma }}}.
\end{align}
Since the righthand side of (\ref{r14}) is minimized when $\sigma=\ln \left( {1 + \frac{r}{{{\mu _X}}}} \right)$, we have
\begin{align}
\label{r15}
{\rm P}\left( {X - {\mu _X} \ge r} \right) < {e^{r - \left( {r + {\mu _X}} \right)\ln \left( {1 + \frac{r}{{{\mu _X}}}} \right)}}.
\end{align}
Taking $r=\delta \mu_X$ in (\ref{r15}), (\ref{r8}) directly follows.
\end{IEEEproof}

With Lemma 1, following the arguments for (12) in \cite{Panli2012}, we have the achievable throughput from SUs to neighboring BSs/CR routers (i.e., the first step) as
\begin{align}
\label{r6}
\xi_1=\Omega(n^{b-1}),
\end{align}
where $g\left( n \right) = \Omega \left( {h\left( n \right)} \right)$ means $\mathop {\lim \inf }\limits_{n \to \infty } \left| {\frac{{g\left( n \right)}}{{h\left( n \right)}}} \right| > 0$.

\begin{figure}[!t]
 \begin{center}
  \includegraphics[width=3in]{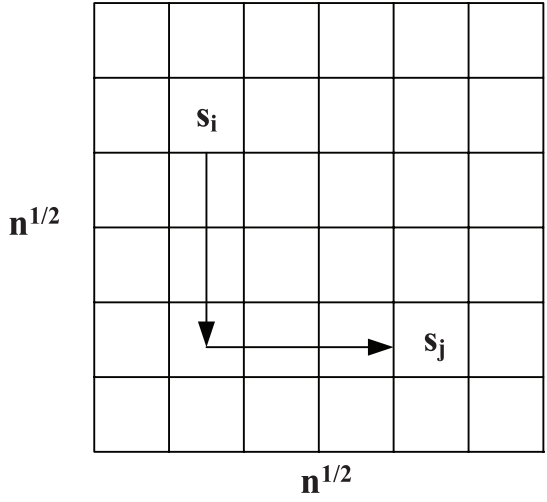}
  \end{center}
  \begin{center}
   \parbox{8cm}{\caption{The adopted routing protocol for data delivery between CR routers and BSs.}  \label{figrouting}}
  \end{center}
  \end{figure}

For the data delivery from CR routers to targeted BSs (i.e., data delivery in step 2), we employ the routing strategy shown in Fig. \ref{figrouting} where $s_i$ represents subsquare where the source resides in and $s_j$ is the subsquare where the targeted BS resides in. As shown in Fig. \ref{figrouting}, packets from the source are first relayed along those subsquares which have the same $x$-coordinate as $s_i$ until they arrive at a subsquare which has the same $y$-coordinate as $s_j$. Then, the packets are relayed along the subsquares which have the same $y$-coordinates until they arrive at $s_j$. Following above assumptions, for each subsquare containing a BS, the flow initiated from this subsquare will not be routed to other subsquares. With Lemma 5 in \cite{Panli2012}, the total number of flows which will be targeted to another subsquare is at most $\varsigma=\min \left\{ {n,2n\left( {1 - {n^{d - b}}} \right)} \right\}$. When $d=b$, i.e., all the nodes deployed by the SSP are BSs, there will be no flows targeted to another subsquare, and the achievable throughput of SUs are determined by that in the first step, $\Omega(n^{b-1})$. When $d<b$, there will be at most $\varsigma$ flows targeted to another square. In this case, the achievable throughput in the second step will be limited by what each subsquare can offer. As proved in \cite{Panli2012}, each subsquare can transmit at a rate of $c_3W$, were $c_3$ is a deterministic positive constant. Then, the achievable throughput is limited by the number of flows to be handled by each subsquare. Denote the number of flows handled by the $i$th subsquare as $\varphi$, the number of source nodes located in the subsquares with the same $x$-coordinated as the $i$th subsquare as $C_x$, and the number of BSs located in the subsquares with the same $y$-coordinated as the $i$th subsquare as $C_y$. Denote the maximum number of flows that a BS can handle as $\overline \varsigma$. We have
\begin{align}
\label{r7}
\varphi \le C_x+\overline \varsigma C_y.
\end{align}

For $\overline \varsigma$, we have the following Lemma.

\begin{lemma}
With high probability, each BS will serve as the destination for at most ${{2\varsigma } \mathord{\left/
 {\vphantom {{2\varsigma } {{n^d}}}} \right.
 \kern-\nulldelimiterspace} {{n^d}}}$ flows, i.e., $\overline \varsigma  {{ = 2\varsigma } \mathord{\left/
 {\vphantom {{ = 2\varsigma } {{n^d}}}} \right.
 \kern-\nulldelimiterspace} {{n^d}}}$.
\end{lemma}

\begin{IEEEproof}
Let ${\overline \varsigma  _i}$ be the number of flows which have the $i$th BS as their destinations. The mean of ${\overline \varsigma  _i}$ is ${\rm E}\left[ {{{\bar \varsigma }_i}} \right] = {\varsigma  \mathord{\left/
 {\vphantom {\varsigma  {{n^d}}}} \right.
 \kern-\nulldelimiterspace} {{n^d}}}$. Then, according to Lemma 1, we have
\begin{align}
\label{r22}
{\rm P}\left( {{{\bar \varsigma }_i} \ge 2{\varsigma  \mathord{\left/
 {\vphantom {\varsigma  {{n^d}}}} \right.
 \kern-\nulldelimiterspace} {{n^d}}}} \right) < {e^{ - {\varsigma  \mathord{\left/
 {\vphantom {\varsigma  {{n^d}}}} \right.
 \kern-\nulldelimiterspace} {{n^d}}}f\left( 1 \right)}}.
\end{align}

Since $f(1)>0$, ${\varsigma  \mathord{\left/
 {\vphantom {\varsigma  {{n^d}}}} \right.
 \kern-\nulldelimiterspace} {{n^d}}} = \min \left\{ {{n^{1 - d}},2{n^{1 - d}}\left( {1 - {n^{d - b}}} \right)} \right\}$, and $1-d>0$, ${\rm P}\left( {{{\bar \varsigma }_i} \ge 2{\varsigma  \mathord{\left/
 {\vphantom {\varsigma  {{n^d}}}} \right.
 \kern-\nulldelimiterspace} {{n^d}}}} \right) \to 0$ when $n\to \infty$. By definition, $\overline \varsigma   = \mathop {\max }\limits_i \left\{ {{{\bar \varsigma }_i}} \right\}$. Then, it follows
\begin{align}
\label{r23}
 {\rm P}\left( {\bar \varsigma  \le 2{\varsigma  \mathord{\left/
 {\vphantom {\varsigma  {{n^d}}}} \right.
 \kern-\nulldelimiterspace} {{n^d}}}} \right) =& 1 - {\rm P}\left( {\mathop {\max }\limits_i \left\{ {{{\bar \varsigma }_i}} \right\} > 2{\varsigma  \mathord{\left/
 {\vphantom {\varsigma  {{n^d}}}} \right.
 \kern-\nulldelimiterspace} {{n^d}}}} \right) \\ \nonumber
  \ge& 1 - \sum\limits_{i = 1}^{{n^d}} {{\rm P}\left( {{{\bar \varsigma }_i} \ge 2{\varsigma  \mathord{\left/
 {\vphantom {\varsigma  {{n^d}}}} \right.
 \kern-\nulldelimiterspace} {{n^d}}}} \right)}  > 1 - {n^d}{e^{ - {\varsigma  \mathord{\left/
 {\vphantom {\varsigma  {{n^d}}}} \right.
 \kern-\nulldelimiterspace} {{n^d}}}f\left( 1 \right)}}.
\end{align}
Clearly, ${n^d}{e^{ - {\varsigma  \mathord{\left/
 {\vphantom {\varsigma  {{n^d}}}} \right.
 \kern-\nulldelimiterspace} {{n^d}}}f\left( 1 \right)}} \to 0$ when $n \to \infty$. Thus, ${\rm P}\left( {\bar \varsigma  \le 2{\varsigma  \mathord{\left/
 {\vphantom {\varsigma  {{n^d}}}} \right.
 \kern-\nulldelimiterspace} {{n^d}}}} \right) \to 1$ as $n \to \infty$, which completes the proof.
\end{IEEEproof}

As for $C_x$ and $C_y$, we have the following Lemma.

\begin{lemma}
With high probability, for each subsquare, we have
\begin{enumerate}
  \item \textit{When} $0<\frac{b}{2}<d<1$, $C_x=\min \left\{ {2{n^{\frac{3}{2} - b}} - 2{n^{1 + d - \frac{{3b}}{2}}},2{n^{1 - \frac{b}{2}}}} \right\}$. \textit{When} $0<d \le \frac{b}{2}<1$, $C_x=\min \left\{ {2{n^{\frac{3}{2} - b}} - 2{c_2}{n^{1 - b}},2{n^{1 - \frac{b}{2}}}} \right\}$.
  \item \textit{When} $0<\frac{b}{2}<d<1$, $C_y=n^{d-\frac{b}{2}}$. \textit{When} $0<d \le \frac{b}{2}<1$, $C_y=c_1$.
\end{enumerate}
\end{lemma}

\begin{IEEEproof}
When $0<\frac{b}{2}<d<1$, there are $n^{d-\frac{b}{2}}$ subsquares sharing the same $x$-coordinate and contain a BS. With Lemma 5 in \cite{Panli2012}, the number of flows which are located in the subsquares with the same $x$-coordinate and target to BSs in other subsquares is at most $2{n^{1 - b}}\left( {{n^{\frac{1}{2}}} - {n^{d - \frac{b}{2}}}} \right)$. Following the same procedure as shown in Lemma 2, we can prove that the number of flows which are located in the subsquares with the same $x$-coordinate is at most $2{n^{1 - \frac{b}{2}}}$. Thus, it follows ${C_x} = \min \left\{ {2{n^{\frac{3}{2} - b}} - 2{n^{1 + d - \frac{{3b}}{2}}},2{n^{1 - \frac{b}{2}}}} \right\}$. Similarly, we can prove when $0<d \le \frac{b}{2}<1$, ${C_x} = \min \left\{ {2{n^{\frac{3}{2} - b}} - 2{c_2}{n^{1 - b}},2{n^{1 - \frac{b}{2}}}} \right\}$.

Noticing that, when $0<\frac{b}{2}<d<1$, there are $n^{d-\frac{b}{2}}$ subsquares sharing the same $y$-coordinate and contain a BS, it follows $C_y=n^{d-\frac{b}{2}}$. Similarly, when $0<d \le \frac{b}{2}<1$, we have $C_y=c_1$.
\end{IEEEproof}

From Lemma 2 and Lemma 3, we can obtain
\begin{align}
\label{r24}
\varphi  \le \left\{ {\begin{array}{*{20}{c}}
{\left\{ {\begin{array}{*{20}{c}}
{\min \left\{ {2{n^{\frac{3}{2} - b}} - 2{n^{1 + d - \frac{{3b}}{2}}},2{n^{1 - \frac{b}{2}}}} \right\}}\\
{ + \frac{{2\varsigma }}{{{n^d}}}{n^{d - \frac{b}{2}}}}
\end{array}} \right\}}&{0 < \frac{b}{2} < d < 1}\\
{\left\{ {\begin{array}{*{20}{c}}
{\min \left\{ {2{n^{\frac{3}{2} - b}} - 2{c_2}{n^{1 - b}},2{n^{1 - \frac{b}{2}}}} \right\}}\\
{ + {c_1}\frac{{2\varsigma }}{{{n^d}}}}
\end{array}} \right\}}&{0 < d \le \frac{b}{2} < 1}
\end{array}} \right.
\end{align}

Thus, the achievable throughput for each flow in step 2 is
\begin{align}
\label{r25}
{\xi _2} = \frac{{{c_3}W}}{\varphi } = \left\{ {\begin{array}{*{20}{c}}
   {\Omega \left( {W{n^{\frac{b}{2} - 1}}} \right)} & {0 < \frac{b}{2} < d < 1}  \\
   {\Omega \left( {W{n^{d - 1}}} \right)} & {0 < d \le \frac{b}{2} < 1}  \\
\end{array}} \right.
\end{align}

With (\ref{r6}) and (\ref{r25}), the achievable throughput of individual SUs can be derived as shown in (\ref{r26}).

\begin{thebibliography}{}
{
    \bibitem{Cisco2016}
    Cisco, ``Cisco visual networking index: global mobile data traffic forecast,'' White Paper, Feb. 2016.
    \bibitem{Bazelon2015}
    C. Bazelon, G. McHenry, ``Substantial licensed spectrum deficit (2015-2019): updating the FCC¡¯s mobile data demand projections,'' Jun. 2015.
    \bibitem{Chen2016}
    Y. Chen, and H.-S. Oh, ``A survey of measurement-based spectrum occupancy modeling for cognitive radios,'' {\em IEEE Commun. Surveys Tuts.}, vol. 18, no. 1, pp. 848--859, Firstquarter 2016.
    \bibitem{McHenry2006}
    M. A. McHenry, P. A. Tenhula, D. McCloskey, D. A. Roberson, and C. S. Hood, ``Chicago spectrum occupancy measurements and analysis and a long-term studies proposal,'' in {\em Proc. TAPAS 2006}, Boston, MA, Aug. 2006.
    \bibitem{Haykin2005}
    S. Haykin, ``Cognitive radio: brain-empowered wireless communications,'' {\em IEEE J. Sel. Areas Commun.}, vol. 23, no. 2, pp. 201--220, Feb. 2005.
    \bibitem{Zhang20141}
    N. Zhang, H. Liang, N. Cheng, Y. Tang, J. W. Mark, and X. Shen, ``Dynamic spectrum access in multi-channel cognitive radio networks,'' {\em IEEE J. Sel. Area Commun.}, vol. 32, no. 11, pp. 2053--2064, Nov. 2014.
    \bibitem{Ren2016}
    J. Ren, Y. Zhang, N. Zhang, D. Zhang, and X. Shen, ``Dynamic channel access to improve energy efficiency in cognitive radio sensor networks,'' {\em IEEE Trans. Wireless Commun.}, vol. 15, no. 5, pp. 3143--3156, May 2016.
    \bibitem{Simeone2008}
    O. Simeone, I. Stanojev, S. Savazzi, Y. Bar-Ness, U. Spagnolini, and R. Pickholtz, ``Spectrum leasing to cooperating secondary ad hoc networks,'' {\em IEEE J. Sel. Areas Commun.}, vol. 26, no. 1, pp. 203--213, Jan. 2008.
    \bibitem{Cao2013}
    B. Cao, J. W. Mark, Q. Zhang, R. Lu, X. Lin, and X. Shen, ``On optimal communication strategies for cooperative cognitive radio networking,'' in {\em Proc. 2013 IEEE INFOCOM}, Turin, Italy, pp. 1726--1734, Apr. 2013.
    \bibitem{Jing2015}
    T. Jing, S. Zhu, H. Li, X. Xing, X. Cheng, Y. Huo, R. Bie, and T. Znati, ``Cooperative relay selection in cognitive radio networks,'' {\em IEEE Trans. Veh. Technol.}, vol. 64, no. 5, pp. 1872--1881, May 2015.
    \bibitem{Duan2014}
    L. Duan, L. Gao, and J. Huang, ``Cooperative spectrum sharing: a contract-based approach,'' {\em IEEE Trans. Mobile Comput.}, vol. 13, no. 1, pp. 174--187, Jan. 2014.
    \bibitem{Feng2014}
    X. Feng, G. Sun, X. Gan, F. Yang, X. Tian, X. Wang, and M. Guizani, ``Cooperative spectrum sharing in cognitive radio networks: a distributed matching approach,'' {\em IEEE Trans. Commun.}, vol. 62, no. 8, pp. 2651--2664, Aug. 2014.
    \bibitem{Li2013}
    W. Li, X. Cheng, T. Jing, X. Xing, ``Cooperative multi-hop relaying via network formation games in cognitive radio networks,'' in {\em Proc. 2013 IEEE INFOCOM}, Turin, Italy, pp. 971--979, Apr. 2013.
    \bibitem{Zhang2013}
    N. Zhang, N. Lu, N. Cheng, J. W. Mark, X. Shen, ``Cooperative spectrum access towards secure information transfer for CRNs,'' {\em IEEE J. Sel. Areas Commun.}, vol. 31, no. 11, pp. 2453--2464, Nov. 2013.
    \bibitem{Long2014}
    Y. Long, H. Li, H. Yue, M. Pan, and Y. Fang, ``SUM: spectrum utilization maximization in energy-constrained cooperative cognitive radio networks,'' {\em IEEE J. Sel. Areas Commun.}, vol. 32, no. 11, pp. 2105 - 2116, Nov. 2014.
    \bibitem{Zhang2014}
    N. Zhang, N. Cheng, N. Lu, H. Zhou, J. W. Mark, X. Shen, ``Risk-aware cooperative spectrum access for multi-channel cognitive radio networks,'' {\em IEEE J. Sel. Areas Commun.}, vol. 32, no. 3, pp. 516--527, Mar. 2014.
    \bibitem{Garcia2004}
    A. Leon-Garcia, I. Widjaja, {\em Communication Networks: Fundamental Concepts and Key Architectures (Second Edition)}. McGraw-Hill Higher Education, 2004.
    \bibitem{Thilina2015}
    K. M. Thilina, E. Hossain, and D. I. Kim, ``DCCC-MAC: a dynamic common control channel-based MAC protocol for cellular cognitive radio networks,'' {\em IEEE Trans. Veh. Technol.}, vol. 65, no. 5, pp. 3597--3613, May 2016.
    \bibitem{Anamalamudi2016}
    S. Anamalamudi, M. Jin, ``Energy-efficient hybrid CCC-based MAC protocol for cognitive radio ad hoc networks,'' {\em IEEE Syst. J.}, vol. 10, no. 1, pp. 358--369, Mar. 2016.
    \bibitem{Pan2012}
    M. Pan, C. Zhang, P. Li, and Y. Fang, ``Spectrum harvesting and sharing in multi-hop CRNs under uncertain spectrum supply,'' {\em IEEE J. Sel. Areas Commun.}, vol. 30, no. 2, pp. 369--378, Feb. 2012.
    \bibitem{Grndalen2011}
    O. Gr{\o}ndalen, M. L{\"a}hteenoja,  P. Gr{\o}nsund, ``Evaluation of business cases for a cognitive radio network based on wireless sensor network,'' in {\em Proc. 2011 IEEE Symposium on New Frontiers in Dynamic Spectrum Access Networks (DySPAN 2011)}, Aachen, Germany, May 2011.
    \bibitem{Cammarano2015}
    A. Cammarano, F. Lo Presti, G. Maselli, L. Pescosolido, C. Petrioli, ``Throughput-optimal cross-layer design for cognitive radio ad hoc networks,'' {\em IEEE Trans. Parallel Distrib. Syst.}, vol. 26, no. 9, pp. 2599--2609, Sept. 2015.
    \bibitem{Fodor2009}
    V. Fodor, I. Glaropoulos, L. Pescosolido, ``Detecting low-power primary signals via distributed sensing to support opportunistic spectrum access,'' in {\em Proc. 2009 IEEE International Conference on Communications (ICC 2009)}, Dresden, Germany, June 2009.
    \bibitem{Li2010}
    H. Li, Y. Cheng, C. Zhou, P. Wan, ``Multi-dimensional conflict graph based computing for optimal capacity in MR-MC wireless networks,'' in {\em Proc. IEEE ICDCS}, Genoa, Italy, pp. 774--783, Jun. 2010.
    \bibitem{Yuan2013}
    X. Yuan,  Y. Shi, Y. T. Hou, W. Lou, and S. Kompella, ``UPS: a united cooperative paradigm for primary and secondary networks,'' in {\em Proc. IEEE MASS}, Hangzhou, China, Oct. 14-16, 2013.
    \bibitem{Yuan2017}
    X. Yuan, Y. Shi, X. Qin, Y. T. Hou, W. Lou, S. Kompella, S.F. Midkiff, and J. H. Reed, ``Beyond overlay: reaping mutual benefits for primary and secondary networks through node-level cooperation,'' {\em IEEE Trans. on Mobile Comput.}, vol.16, no. 1, pp. 2-15, Jan. 2017.
    \bibitem{Nadkar2011}
    T. Nadkar, V. Thumar, G. Shenoy, A. Mehta, U. B. Desai, and S. N. Merchant,``A cross-layer framework for symbiotic relaying in cognitive radio networks,'' in {\em Proc. 2011 IEEE International Symposium on New Frontiers in Dynamic Spectrum Access Networks (DySPAN 2011)}, Aachen, Germany, May 2011.
    \bibitem{Yue2013}
    H. Yue, M. Pan, Y. Fang and S. Glisic, ``Spectrum and energy efficient relay station placement in cognitive radio networks,'' {\em IEEE J. Sel. Areas Commun.}, vol. 31, no. 5, May 2013.
    \bibitem{Pan2014}
    M. Pan, P. Li, Y. Song, Y. Fang, P. Lin and S. Glisic, ``When spectrum meets clouds: optimal session based spectrum trading under spectrum uncertainty,'' {\em IEEE J. Sel. Areas Commun.}, vol. 32, no. 3, pp. 615--627, Mar. 2014.
    \bibitem{Li2016}
    X. Li, H. Ding, M. Pan, Y. Sun and Y. Fang, ``Users first: service-oriented spectrum auction with a two-tier framework support,'' {\em IEEE J. Sel. Areas Commun.}, accepted for publication, Sept. 2016.
    \bibitem{Jain2003}
    K. Jain, J. Padhye, V. N. Padmanabhan, and L. Qiu, ``Impact of interference on multi-hop wireless network performance,'' in {\em Proc. Mobicom¡¯03}, San Diego, USA, pp. 66¨C80, Sept. 2003.
    \bibitem{Shi2009}
    Y. Shi, Y. T. Hou, and S. Kompella, ``How to correctly use the protocol interference model for multi-hop wireless networks,'' in {\em Proc. ACM Int. Symp. Mobile Ad Hoc Netw. Comput. (MobiHoc)}, May 2009.
    \bibitem{Rebecchi2015}
    F. Rebecchi, M. D. d. Amorim, V. Conan, A. Passarella, R. Bruno, and M. Conti, ``Data offloading techniques in cellular networks: a survey,'' {\em IEEE Commun. Surveys Tuts.}, vol. 17, no. 2, pp. 580--603, Secondquarter 2015.
    \bibitem{Panli2012}
    P. Li, and Y. Fang, ``On the throughput capacity of heterogeneous wireless networks,'' {\em IEEE Trans. Mobile Comput.}, vol. 11, no. 12, Dec. 2012.

}
\end{thebibliography}
\end{document}